\documentclass[]{aa}

\usepackage[utf8]{inputenc}
\usepackage{graphicx}
\usepackage[varg]{txfonts}
\usepackage{subcaption}
\usepackage{multirow}

\bibpunct{(}{)}{;}{a}{}{,}

\begin{document}
\title{The ultra-deep H\,{\sc i} radial profiles of late-type galaxies from MHONGOOSE}

\author{S. Veronese\inst{\ref{astron},\ref{kapteyn},\ref{mpifr}} \and W. J. G. de Blok\inst{\ref{astron},\ref{kapteyn},\ref{uct}} \and E. Brinks \inst{\ref{car}} \and D. Kleiner\inst{\ref{astron}} \and S. Kurapati\inst{\ref{astron}} \and J. Healy\inst{\ref{manchester},\ref{uct}} \and F. M. Maccagni\inst{\ref{inafca}} }

\institute{Netherlands Institute for Radio Astronomy (ASTRON), Oude Hoogeveensedijk 4, 7991 PD Dwingeloo, The Netherlands, \email{veronese.astro@gmail.com}\label{astron}
\and Kapteyn Astronomical Institute, University of Groningen, PO Box 800, 9700 AV Groningen, The Netherlands\label{kapteyn}
\and Max Planck Institute for Radio Astronomy, Auf dem Hügel 69, 53121 Bonn, Germany\label{mpifr}
\and Department of Astronomy, University of Cape Town, Private Bag X3, 7701 Rondebosch, South Africa\label{uct}
\and Centre for Astrophysics Research, University of Hertfordshire, College Lane, Hatfield AL10 9AB, United Kingdom\label{car}
\and Jodrell Bank Centre for Astrophysics, School of Physics and Astronomy, University of Manchester, Oxford Road, Manchester M13 9PL, UK\label{manchester}
\and INAF – Osservatorio Astronomico di Cagliari, Via della Scienza 5, 09047, Selargius, CA, Italy\label{inafca}}

\date{Received 3 March 2026 / Accepted 16 June 2026}

\abstract{Galaxy discs have a finite extent, yet how and where their neutral atomic hydrogen (H\,{\sc i}) components end is not fully understood. The existence of a break in the H\,{\sc i} disc at column densities of $\sim10^{19}$ cm$^{-2}$ has been debated since early 21-cm observations of the spiral galaxy NGC 3198. We present the H\,{\sc i} radial profiles of 16 star-forming, late-type galaxies from the MeerKAT H\,{\sc i} Observations of Nearby Galactic Objects: Observing Southern Emitters (MHONGOOSE) survey spanning a range of H\,{\sc i} mass from $10^{7}$ M$_\odot$ to $10^{10}$ M$_\odot$. We probed via spectral stacking their H\,{\sc i} discs down to inclination-corrected column densities of a few times $10^{17}$ cm$^{-2}$ at kpc resolution. The H\,{\sc i} radial profiles of high-mass (M$_{\rm {HI}}>10^{9}$ M$_\odot$) galaxies are characterised by a inner plateau, followed by a knee at column densities of $\sim5\times10^{20}$ cm$^{-2}$, but no edges are unambiguously identified. The H\,{\sc i} radial profiles of low-mass (M$_{\rm {HI}}>10^{9}$ M$_\odot$) galaxies shows a steeper decline and also no edges. We found that the profiles are self-similar when normalised to the radius at which a mass surface density of 0.01 M$_\odot$ pc$^{-2}$ is reached, in agreement with recent literature results, but we also found a possible correlation between the normalisation radius and the environment, suggesting that environment contributes to the shaping of the H\,{\sc i} distribution in galaxies. The emerging picture is that the diverse morphology of the H\,{\sc i} radial profiles is difficult to interpret, and future studies with a larger sample are necessary for quantifying the contribution of internal and external processes acting at different levels for different galaxies.}

\maketitle

\section{Introduction}\label{sec:intro}
Galaxy discs probed by stars and gas have a finite extent. It is well known that the radial surface brightness distribution of stellar discs is exponential \citep{freeman70,degrijs96} over a wide range of galaxy luminosities \citep{blok95,dejong96,kregel02,pohlen06}, yet it is debated whether they exhibit an outer edge or a break in the surface brightness (\citealt{kregel02,kruit07,kruit11,comeron12,lombilla19}; see also the work by \citealt{minniti11} on the stellar edge of the Milky Way). The radial column density distribution of the gaseous disc in the outskirts of galaxies is also not well understood. The first observational indication of a steep edge in the gas distribution was presented by \citet{gorkom91} studying the neutral atomic hydrogen (H\,{\sc i}) emission from the spiral galaxy NGC 3198. It was shown that on the NE side of the galaxy the H\,{\sc i} column density suddenly drops toward large radii, going from $\sim10^{20}$ cm$^{-2}$ to $\sim10^{18}$ cm$^{-2}$ in few kpc. Subsequent H\,{\sc i} observations of other galaxies down to column densities of a few times $10^{18}$ cm$^{-2}$ provided more ambiguous results, with the outer parts of galaxies showing extended discs \citep{braun04,westmeier05,wolfe13,wolfe16,pisano14,ianjam18}.
\\\indent\citet{gorkom91} pointed out that interpreting the observed edge in NGC 3198 is non-trivial, because of the strong lopsidedness of the galaxy. The steep decline could represent a physical edge, or it could be the consequence of the photoionisation of H\,{\sc i} caused by an incident cosmic ionising background (CIB; e.g., \citealt{gorkom91,maloney93,dove94,weymann01,corbelli02,popping09,haardt12,madau15,bland17}). Currently, there is no direct measurement of the CIB intensity, but only indirect estimates \citep{weymann01,haardt12,bland17}. Having a precise measurement is fundamental to our understanding of galaxy evolution, as it sets the condition for the ionised-to-neutral gas ratio \citep{faucher09,rahmati13,faucher15}, which impacts the yet-to-be-understood cosmic gas accretion onto galaxies (see the reviews by \citealt{putman12,tumlinson17,peroux20,faucher23}).
\\\indent The H\,{\sc i} radial profile, which describes the average H\,{\sc i} distribution at given radii from the galaxy centre, can thus provide an empirical baseline to constrain the CIB intensity. Indeed, the CIB intensity is a parameter of the photoionisation models used to describe the observed shapes. However, the morphology of the H\,{\sc i} radial profile could also depend on other quantities, such as gas distribution and kinematics, or it could also be that the steep edge is a physical feature (i.e. it marks a decrease in the true mass density of hydrogen).
\\\indent Past studies of H\,{\sc i} radial profiles in a large sample of galaxies were mostly limited by column densities of a few times $10^{19}$ cm$^{-2}$, that is, the column density at which the steep edge is supposed to occur. For instance, using a $3\sigma$ over 16 km s$^{-1}$ definition for the H\,{\sc i} column density sensitivity, the Westerbork HI Survey of Spiral and Irregular Galaxies (WHISP; \citealt{whisp}) on the Westerbork Synthesis Radio Telescope (WSRT; \citealt{wsrt}) reached column densities of a few times $10^{19}$ cm$^{-2}$ at $60''$ angular resolution, The HI Nearby Galaxies Survey (THINGS; \citealt{things}) on the Very Large Array (VLA; \citealt{vla}) reached $\sim10^{19}$ cm$^{-2}$ at $30''$ and the Hydrogen Accretion in LOcal GAlaxieS (HALOGAS; \citealt{halogas}), again on the WSRT, managed to go as low as $5\times10^{18}$ cm$^{-2}$ at $30''$ (see also \citealt{maccagni24b} for a comprehensive description about past H\,{\sc i} surveys and \citealt{ianjam18} about previous studies of THINGS and HALOGAS H\,{\sc i} radial profiles). None of them was able to unambiguously identify edges in the gaseous discs.
\\\indent Along with column density sensitivity, also angular resolution is a limiting factor in determining the presence of breaks. The work by \citet{sardone21} presented H\,{\sc i} radial profiles of 18 nearby galaxies, some of which are also included in this paper, observed with the Green Bank Telescope \citep{gbt}. They probed the gaseous disc down to column densities of $6.3\times10^{17}$ cm$^{-2}$ ($3\sigma$ over 20 km s$^{-1}$) but galaxies were mostly unresolved at the angular resolution of $9'.1$, making it impossible to determine the existence of breaks. Similarly, the recent work by \citet{wang25} probed the H\,{\sc i} radial profile of 35 nearby galaxies with the FAST \citep{fast} telescope down to column densities of $\sim10^{18}$ cm$^{-2}$ ($3\sigma$ over 20 km s$^{-1}$). Again, the angular resolution of $3'.24$ was not sufficient to resolve galaxies with enough resolution elements to detect any break.
\\\indent With the MeerKAT telescope \citep{meerkat}, we now have the observational capabilities to reach in reasonable observing times limiting column densities of $\sim10^{18}$ cm$^{-2}$ at $30''$ and even a few times $10^{17}$ cm$^{-2}$ at $60''$. Indeed, the recent MeerKAT H\,{\sc i} Observations of Nearby Galactic Objects: Observing Southern Emitters (MHONGOOSE, \citealt{mhongoose2}) survey represents a major step forward in the empirical study of H\,{\sc i} in the local universe. MHONGOOSE provides H\,{\sc i} observations of 30 nearby gas-rich dwarf and spiral galaxies with different stellar masses ($4.7\leq\log M_*\leq10.7$ M$_\odot$) and star formation rates ($-2.52\leq\log\text{SFR}\leq0.75$ M$_\odot$ yr$^{-1}$). The survey reaches column densities down to a few times $10^{17}$ cm$^{-2}$, angular resolution from $7''$ to $90''$ and a spectral resolution of 1.4 km s$^{-1}$ in a 1.5$^\circ$-wide field around each target. Hence, the MHONGOOSE sample represents the ideal dataset to improve our understanding of the H\,{\sc i} distribution in the outskirts of galaxies. In this paper, we present the radial profiles of a subsample of MHONGOOSE galaxies and investigate the possible relationship between their properties and the morphology of their H\,{\sc i} radial profile.
\\\indent This paper is organised as follows. In Sect.\ \ref{sec:data} we present how the H\,{\sc i} data have been reduced and the MHONGOOSE subsample has been selected. In Sect.\ \ref{sec:ana} we describe how we derived the H\,{\sc i} radial profiles and in Sect.\ \ref{sec:disc} we discuss the results, focusing on identifying edges and exploring alternative mechanisms driving the shape of H\,{\sc i} profiles. Finally, in Sect.\ \ref{sec:conc} we list the conclusions and briefly provide some future prospects.

\section{Data reduction and sample selection}\label{sec:data}
Each of the 30 MHONGOOSE galaxies was observed for a total of 55 h, spread over 10 observing sessions lasting 5.5 h each. The 10 observation tracks underwent the same calibration procedure through the Containerized Automated Radio Astronomy Calibration (\texttt{CARACal}) pipeline \citep{caracal} as described in \citet{mhongoose2}. \texttt{CARACal} provides a unified platform for standard reduction steps, including data flagging, calibration, continuum subtraction, and spectral line imaging. The H\,{\sc i} cubes were generated via \texttt{WSClean} \citep{wscleana,wscleanb} within the \texttt{CARACal} pipeline. The cleaning of H\,{\sc i} data followed a three-step iterative strategy using the Source Finding Application (\texttt{SoFiA-2}, \citealt{sofia2}) to create masks for cleaning.
\\\indent In this paper, we look for resolved H\,{\sc i} in the outskirts of galaxies. This requires a balance between angular resolution and sensitivity. Therefore, we opt for the $30''$ cubes, obtained by weighting the visibilities with a robust parameter of 1.5 and without any tapering. These data reach an average $3\sigma$ limiting column density sensitivity of a few times $10^{18}$ cm$^{-2}$ over 16 km s$^{-1}$. We used the intensity maps (moment 0) and the intensity-weighted mean velocity fields (moment 1) produced by \texttt{SoFiA-2} within the MHONGOOSE pipeline. For an extensive description of the data reduction procedure and creation of the various maps see \citet{mhongoose2}.
\\\indent As we want to relate any features in the outer H\,{\sc i} profile to the intrinsic density distribution of the H\,{\sc i}, rather than other processes like environmental effects, we excluded MHONGOOSE galaxies that are interacting. The MHONGOOSE sample was initially selected to include only isolated galaxies; however, thanks to the deeper MeerKAT observations we found that two targets (J1103-23 and J1254-10a) are actually close systems (Kleiner et al., in prep). We also excluded edge-on galaxies (inclination $\geq80^\circ$; J0516-37, J1153-28, J1303-17b and J2009-61) because of our approach to determine the radial profile by averaging the column density in annuli. Finally, we did not consider galaxies with irregular kinematics, based on a preliminary visual inspection of the moment 1 map and a proper kinematical model as done in Sect.\ \ref{sec:rings}. These kinematically irregular galaxies are dwarfs (M$_{\rm{HI}}<10^{8.1}$ M$_\odot$) J0049-20, J0320-52, J1321-31, J0429-27, J0454-53, J0135-41 and the more massive J0331-51 and J0546-52. The final sample consists of 16 out of 30 MHONGOOSE targets that we list in Table \ref{table:sample}.

\begin{table*}
    \small
    \caption{Basic properties of the MHONGOOSE galaxies in our sample}
    \label{table:sample}
    \centering
    \begin{tabular}{c c c c c c c c c}
        \hline
        \hline
        HIPASS & Name & $\alpha$ (J2000.0) & $\delta$ (J2000.0) & $D$ & $\log(M_{\rm HI})$ & $i$ & $\log(M_{\rm \star})$ & log(SFR)\\
        & & [h m s] & $[^\circ\ '\ '']$ & [Mpc]& [M$_\odot$] & $[{}^{\circ}]$& [M$_\odot$] & [M$_\odot$ yr$^{-1}$] \\
        \hline
        J0008-34 & ESO349-G031 & 00 08 13.36 & -34 34 42.0 & 3.3  & 7.11 & 35 & 6.13 & -2.52\\
        J0031-22 & ESO473-G024 & 00 31 22.51 & -22 45 57.5 & 7.2  & 7.95 & 57 & 6.84 & -2.22\\
        J0052-31 & NGC0289     & 00 52 42.36 & -31 12 21.0 & 21.5 & 10.35 & 45 & 10.43 & 0.16\\
        J0309-41 & ESO300-G014 & 03 09 37.87 & -41 01 49.7 & 10.9 & 8.89 & 59 & 8.90 & -0.99\\
        J0310-39 & ESO300-G016 & 03 10 10.48 & -40 00 10.5 & 8.0  & 7.80 & 36 & 7.22 & -2.52\\
        J0351-38 & ESO302-G014 & 03 51 40.90 & -38 27 08.0 & 16.8 & 8.90 & 28 & 8.19 & -2.52\\
        J0335-24 & NGC1371     & 03 35 01.34 & -24 55 59.6 & 22.7 & 9.97 & 46 & 10.63 & -0.62\\
        J0419-54 & NGC1566     & 04 20 00.42 & -54 56 16.1 & 17.7 & 10.08 & 37 & 10.68 & 0.61\\
        J0445-59 & NGC1672     & 04 45 42.50 & -59 14 49.9 & 19.4 & 10.29 & 34 & 10.69 & 0.75\\
        J0459-26 & NGC1744     & 04 59 57.80 & -26 01 20.0 & 9.3  & 9.54 & 57 & 9.18 & -0.78\\
        J1106-14 & KKS2000-23  & 11 06 12.00 & -14 24 25.7 & 13.9 & 8.74 & 70 & 7.51 & -1.82\\
        J1253-12 & UGCA307     & 12 53 57.29 & -12 06 21.0 & 11.0 & 8.88 & 60 & 8.05 & -1.56\\
        J1318-21 & NGC5068     & 13 18 54.81 & -21 02 20.8 & 5.2  & 9.01 & 28 & 9.36 & -0.54\\
        J1337-28 & ESO444-G084 & 13 37 19.99 & -28 02 42.0 & 4.6  & 8.02 & 40 & 6.69 & -2.52\\
        J2257-41 & NGC7424     & 22 57 18.37 & -41 04 14.1 & 7.9  & 9.60 & 32 & 9.56 & -0.46\\
        J2357-32 & NGC7793     & 23 57 49.83 & -32 35 27.7 & 3.6  & 8.96 & 47 & 9.34 & -0.71\\
	\hline
    \end{tabular}
    \tablefoot{The inclination angles refer to the optical disc. For further details see Table 1 of \citet{mhongoose2}.}
\end{table*}

\section{Data analysis}\label{sec:ana}
We derived the H\,{\sc i} radial profile by averaging the column density in a set of rings (e.g., \citealt{maloney93,ianjam18}). This approach has the advantage of maximising the signal-to-noise ratio (S/N), hence going as deep as the data allow. Following the procedure described in \citet{ianjam18}, we divided each galaxy into concentric elliptical annuli. We set the ring width along the major axis equivalent to the major axis of the beam to ensure that each radial point is independent \citep{begeman87,gentile13,oh15,mancera20,mancera22}. The H\,{\sc i} radial profile is then simply the average column density within each annulus, corrected for the inclination of the ring. The calculation can be done directly on the moment 0 map or on the cube via spectral stacking to further increase the sensitivity of the data and reach larger radii. In the following, we explain how we determine the H\,{\sc i} radial profiles of our sample. We refer the reader to Sect.\ \ref{sec:test} for a thorough discussion about how we tested the methodology against mock and real data and about how we quantified the main sources of uncertainties.

\subsection{Spectral stacking}\label{sec:stack}
\begin{figure}
    \centering
    \includegraphics[width=\hsize]{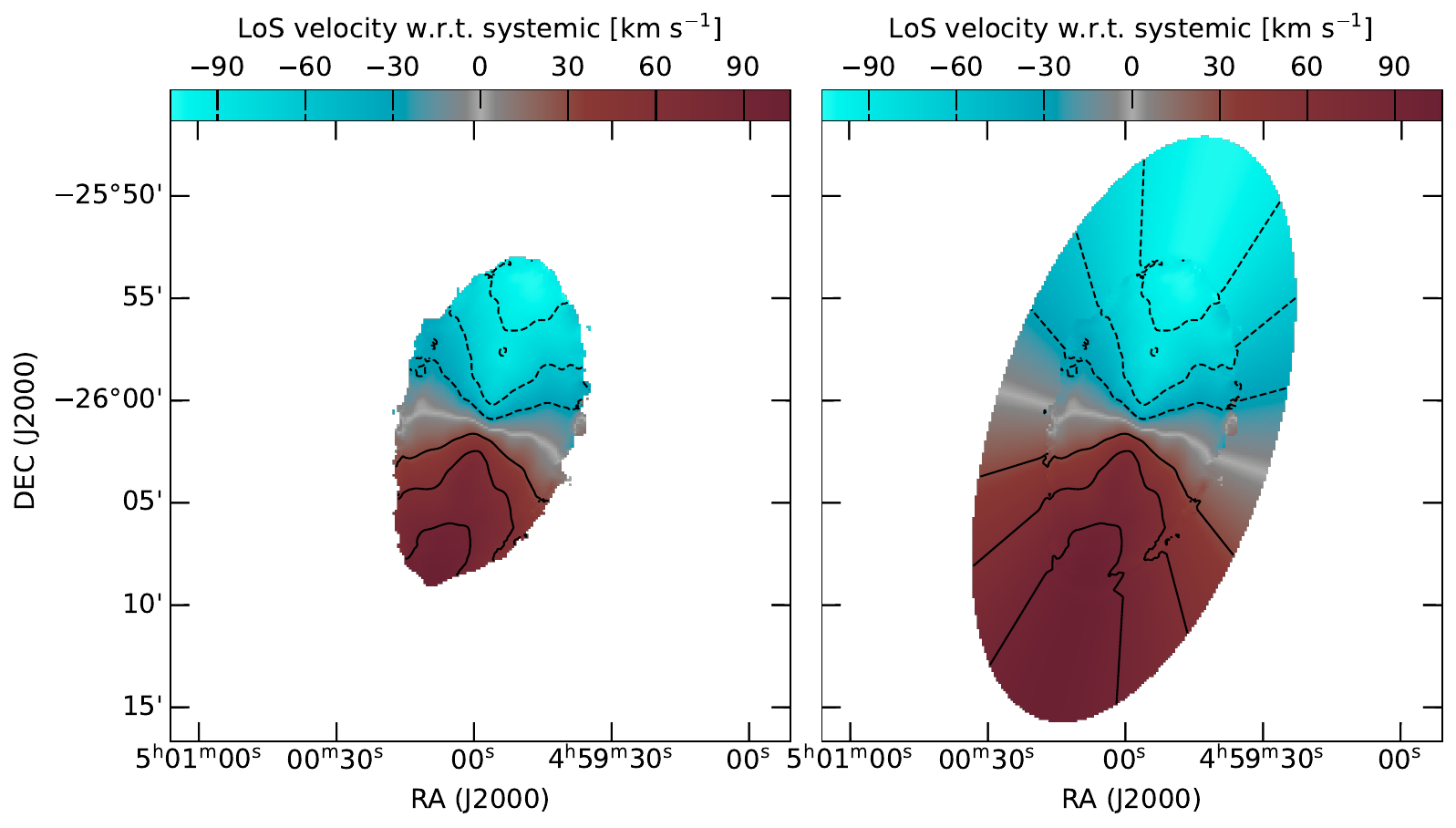}
    \caption{Extension of the J0459-26 velocity field. The left-side panel shows the moment 1 map derived from \texttt{SoFiA-2} as part of the MHONGOOSE pipeline. Dashed contours refer to the approaching side, whereas solid lines correspond to receding velocities. In the right-side panel the extended velocity field is shown which is used to spectrally align the cube for stacking.}
    \label{fig:velfi}
\end{figure}

The H\,{\sc i} radial profile is retrieved directly from the moment 0 map or by spectrally stacking the cube within the rings. Spectral stacking \citep{zwaan00,chengalur01,fabello11,delhaize13,kanekar16,healy21a,amiri23} consists of extracting spectra along $N$ different lines of sight (LoS), aligning them to a reference velocity \citep{khandai11,maddox13,neumann23} and co-adding them so that the flux density in a given channel $k$ of the stacked spectrum ($F_{s,k}$) is
\begin{equation}
    F_{s,k}=\frac{\sum^N_{i=1}F_{i,k}w_i}{\sum^N_{i=1}w_i}
,\end{equation}
where $F_{i,k}$ is the flux of the spectrum $i$ in the channel $k$ and $w_i$ is a weight. We chose $w_i=\frac{1}{\sigma_i^2}$, where $\sigma_i^2$ is the noise in the $i$th spectrum, to give less weight to noisy spectra (e.g.\ \citealt{fabello11,kanekar16,ianjam18,hiss,healy21a,chowdhury22,apurba23}).
\\\indent The effectiveness of spectral stacking is mainly based on two conditions: Gaussian noise and spectral alignment. The former guarantees that the noise decreases as $\sqrt{N}$, the latter maximises the S/N of the stacked spectrum. \citet{veronese25} have shown that the MHONGOOSE noise can be assumed to be Gaussian. Therefore, the first condition holds. For the spectral alignment we used an extended velocity field, obtained by combining the observed moment 1 map with an extension to larger radii. This extension was computed by assuming that the orientation of the gas disc on the sky and the rotational velocity are constant and equal to the values of the outermost annulus. Although constant rotation velocity extension is a common choice in the literature (e.g. \citealt{ianjam18}), assuming that the geometry is also constant might be an oversimplification, as H\,{\sc i} discs are known to warp in the outermost regions \citep{garcia02}. We will briefly discuss in Sect.\ \ref{sec:error} how we quantified the impact of these assumptions. In Fig.\ \ref{fig:velfi} we show the result of the velocity field extension for the galaxy J0459-26\footnote{The velocity field colour map look-up table throughout this paper is taken from \citet{canvas}.}. This extended velocity field is used to spectrally align the cube. The alignment consists of shifting each spectrum along the spectral axis so that its central channel corresponds to the value of the same LoS in the extended velocity field.
\\\indent A line finder \citep{veronese25} was used to detect emission in the stacked spectra and quantify its flux. The line finder, which is based on the \texttt{SoFiA-2} algorithm \citep{sofia,sofia2}, applies an initial $\sigma$-clipping and then smooths each stacked spectrum with a boxcar kernel, applying a $\sigma$-clipping again after each iteration. Here we choose a kernel equal to 15 channels (21 km s$^{-1}$) and opt for a $\sigma$-threshold of 3. Detections within 3 channels (4.2 km s$^{-1}$) are grouped together into the same source. The resulting source is deemed reliable if it spans more than 7 channels (9.8 km s$^{-1}$), that is the typical dispersion of the H\,{\sc i} line \citep{leroy08,ianjam12}. We visually inspected the stacked spectra to confirm the detections of the line finder. The total flux of a line is finally calculated as the sum of the flux in each detected channel.

\begin{table}
    \caption{Kinematical centre, position and inclination angle of the rings used to compute the H\,{\sc i} radial profiles.}
    \label{table:angles}
    \centering
    \begin{tabular}{c | c | c | c}
    \hline\hline
        Galaxy & Centre (J2000) & $i$ & $\varphi $\\
        & [h m s] , $[^\circ\ '\ '']$ & $[{}^{\circ}]$ & $[{}^{\circ}]$\\
        \hline
        J0008-34 & 00 08 11.66, -34 35 03.0 & $42\pm1$ & $219\pm3$ \\
        J0031-22 & 00 31 24.03, -22 46 18.5 & $45\pm2$ & $173\pm8$ \\
        J0052-31 & 00 52 43.45, -31 12 35.0 & $55\pm2$ & $130\pm8$ \\
        J0309-41 & 03 09 39.73, -41 02 10.7 & $59\pm1$ & $170\pm7$ \\
        J0310-39 & 03 10 12.31, -40 00 24.5 & $44\pm1$ & $190\pm2$ \\
        J0335-24 & 03 35 03.40, -24 56 27.6 & $46\pm1$ & $151\pm8$ \\
        J0351-38 & 03 51 40.90, -38 27 57.0 & $34\pm2$ & $135\pm3$ \\
        J0419-54 & 04 20 02.05, -54 56 30.1 & $40\pm1$ & $223\pm6$ \\
        J0445-59 & 04 45 46.15, -59 15 17.9 & $48\pm2$ & $99\pm15$ \\
        J0459-26 & 04 59 58.84, -26 01 34.0 & $62\pm1$ & $162\pm3$ \\
        J1106-14 & 11 06 13.44, -14 24 46.7 & $59\pm2$ & $238\pm4$ \\
        J1253-12 & 12 53 58.72, -12 06 42.0 & $61\pm2$ & $138\pm4$ \\
        J1318-21 & 13 18 56.31, -21 02 41.8 & $53\pm12$ & $224\pm40$ \\
        J1337-28 & 13 37 21.57, -28 03 03.0 & $47\pm1$ & $70\pm2$ \\
        J2257-41 & 22 57 20.23, -41 04 28.1 & $26\pm2$ & $226\pm5$ \\
        J2357-32 & 23 57 50.94, -32 35 48.7 & $42\pm2$ & $293\pm6$
    \end{tabular}
    \tablefoot{The angles are the medians over all rings as determined from the tilted-ring fit, while the uncertainties are the standard deviation. Values for J0335-24 and J1318-21 were taken from \citet{veronese25b} and \citet{healy24}, respectively.}
\end{table}

\subsection{Rings parametrisation}\label{sec:rings}
The geometry of the rings used to extract the H\,{\sc i} radial profiles of our MHONGOOSE galaxies was determined by performing a 3D fit on the H\,{\sc i} cubes with the \texttt{3DFIT} task of the 3D-Based Analysis of Rotating Objects via Line Observations (\texttt{3D-Barolo}; \citealt{barolo}). This procedure consists of modelling the disc of the galaxy with a set of rings, each with its own rotation centroid ($x_0$, $y_0$), scale height ($z_0$), inclination and position angles ($i$, $\varphi$). The latter defined as the angle the receding major axis makes in the plane of the sky measured from North through East. The velocity of the gas along a LoS in a given ring is computed as \citep{begeman87}
\begin{equation}\label{eq:vlos}
    v_\text{los}=v_\text{sys}+(v_\text{rot}\cos{\theta}+v_\text{rad}\sin{\theta})\sin{i}
,\end{equation}
where $v_\text{sys}$, $v_\text{rad}$, and $v_\text{rot}$ are the systemic, radial, and rotation velocity, respectively, whereas $\theta$ is the azimuthal angle in the plane of the disc, measured with respect to the major axis position angle $\varphi$. The five geometric variables ($x_0$, $y_0$, $z_0$, $i$, $\varphi$), the three kinematic parameters ($v_\text{sys}$, $v_\text{rot}$, $v_\text{rad}$), and the intrinsic velocity dispersion ($\sigma_0$) are all simultaneously fitted for each ring. We followed the strategy described in detail in \citet{diteodoro21}. It consists of a two-stage fit, done on the approaching and receding side separately, as well as on both sides of the galaxy simultaneously. In the first stage, the free parameters are fitted. However, a systematic issue in 3D modelling is that the derived $\varphi$ and $i$ often show the largest scatter \citep{barolo}. Thus, a second stage regularises $\varphi$ and $i$ and repeats the fit on the remaining free parameters. We chose Bezier functions \citep{barolo} for this regularisation, as for most of our galaxies $\varphi$ and $i$ cannot be approximated by a constant or a simple polynomial. For galaxy J0335-24 we used the model presented in \citet{veronese25b}, where the same method was applied, and for galaxy J1318-21 we retrieved $\varphi$ and $i$ from \citet{healy24}.
\\\indent The goodness of the fit was assessed with a visual comparison of the data and the model using diagnostic plots. These include the moment 1 residuals and the position-velocity (\textit{pv}) diagram (see Figs.\ \ref{fig:mom1} and \ref{fig:pv} for a representative example based on the galaxy J0459-26). The $pv$ is a slice across the cube that gives the velocity of the gas as a function of a spatial coordinate, in this case the distance from the centre of the galaxy. Since the velocity of the gas along the LoS is given by Eq.\ (\ref{eq:vlos}), the $pv$ along the major axis ($\theta=0^\circ$, top-left panel of Fig.\ \ref{fig:pv}) traces the rotation curve of the galaxy if the position angle is constant as a function of radius. Instead, along the minor axis ($\theta=90^\circ$, bottom-left panel of Fig.\ \ref{fig:pv}), non-circular motions can be identified via deviations from the systemic velocity. We finally considered the median value of the position and inclination angle of all rings to build the set used to calculate the H\,{\sc i} radial profiles. These values are listed for each galaxy in Table \ref{table:angles}.

\begin{figure}
    \centering
    \includegraphics[width=\hsize]{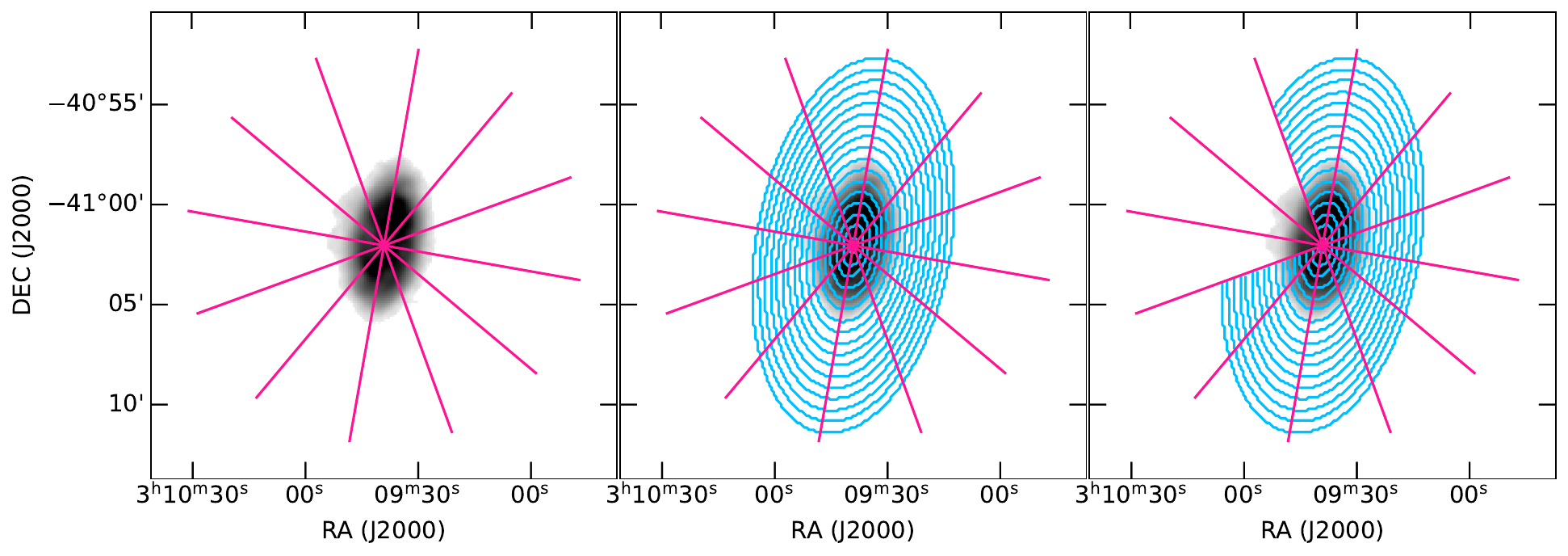}
    \caption{Elliptical regions used to derive the H\,{\sc i} radial profile for J0309-41. In all panels, the moment 0 map of the galaxy is overlaid with the $30^\circ$-wide conical sectors, indicated with magenta lines, used to filter out the disturbances in the disc. In the central panel we overlay also the initial rings (in blue), whereas in the right-side panel we show only the parts of the annuli used to derive the clean H\,{\sc i} radial profile.}
    \label{fig:rings}
\end{figure}

\begin{table}
    \caption{Azimuthal angles, with respect to the position angle of the receding side measured in the plane of the sky, excluded from the computation of the H\,{\sc i} radial profile.}
    \label{table:sector}
    \centering
    \begin{tabular}{c | c c c | c c c | c}
    \hline\hline
        Galaxy & $\theta$\\
        \hline
        J0008-34 & none \\
        J0031-22 & none \\
        J0052-31 & 90$^\circ$-360$^\circ$ \\
        J0309-41 & 210$^\circ$-300$^\circ$ \\
        J0310-39 & none \\
        J0335-24 & 0$^\circ$-150$^\circ$, 270$^\circ$-330$^\circ$\\
        J0351-38 & 120$^\circ$-180$^\circ$\\
        J0419-54 & 0$^\circ$-150$^\circ$, 270$^\circ$-360$^\circ$\\
        J0445-59 & 0$^\circ$-60$^\circ$, 150$^\circ$-300$^\circ$, 330$^\circ$-360$^\circ$\\
        J0459-26 & none \\
        J1106-14 & none \\
        J1253-12 & 90$^\circ$-120$^\circ$, 180$^\circ$-330$^\circ$\\
        J1318-21 & 30$^\circ$-120$^\circ$ \\
        J1337-28 & none \\
        J2257-41 & none \\
        J2357-32 & none \\
    \end{tabular}
\end{table}

\begin{figure}
    \centering
    \includegraphics[width=\hsize]{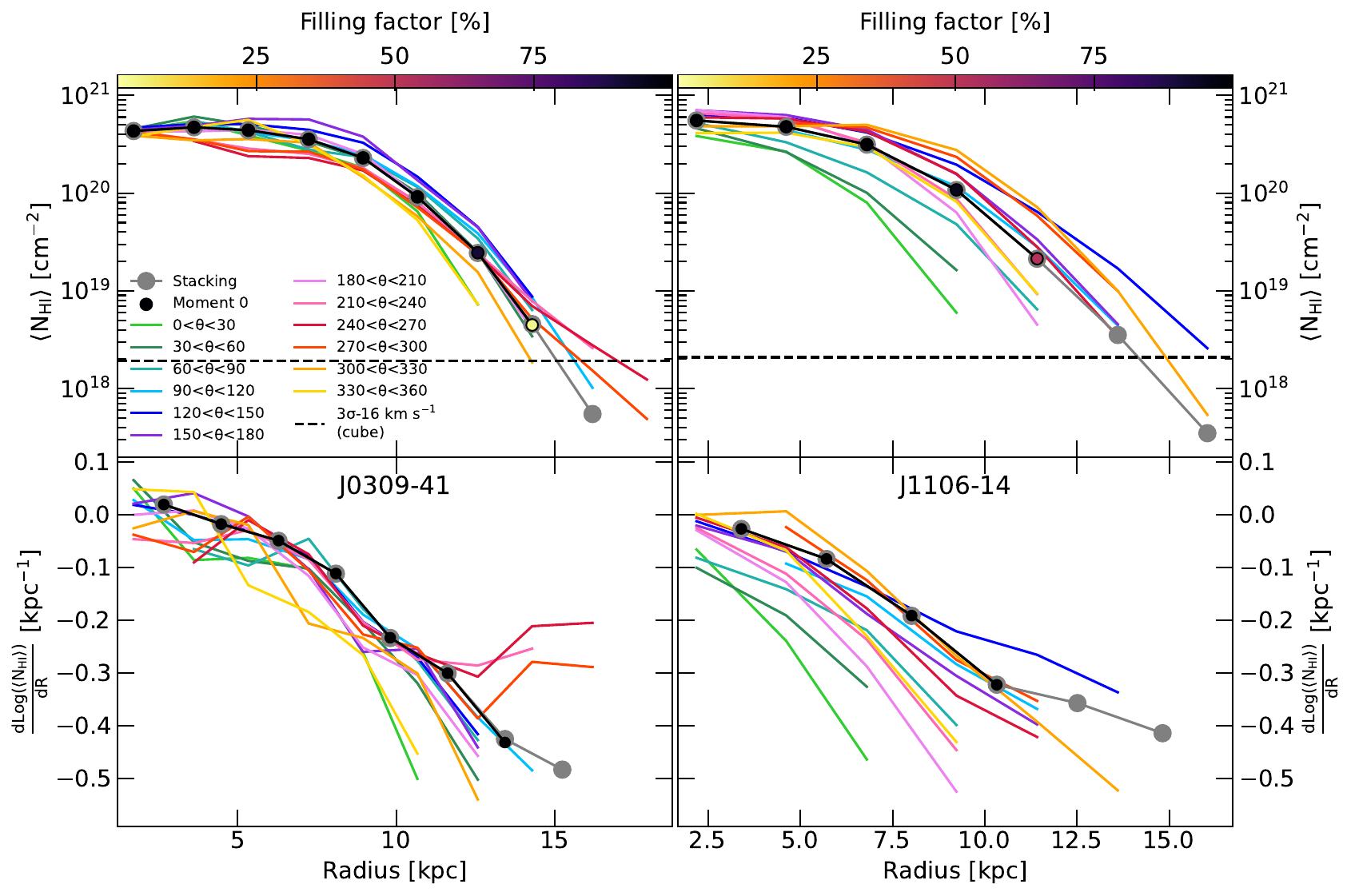}
    \caption{H\,{\sc i} radial profile (top row) and its slope (bottom row) for each azimuthal sector for galaxies J1106-14 (left) and J0309-41 (right). The H\,{\sc i} radial profile derived from the entire galaxy via stacking is reported with gray points. The points colour-coded with the ring filling factor refer to the H\,{\sc i} radial profile derived for the moment 0 map. The coloured lines are the radial profile of each azimuthal sector as recovered via stacking only (see text for further details). The black dashed line is the $3\sigma$ noise level in the cube over 16 km s$^{-1}$.}
    \label{fig:sectors}
\end{figure}

\begin{figure}
    \centering
    \includegraphics[width=\hsize]{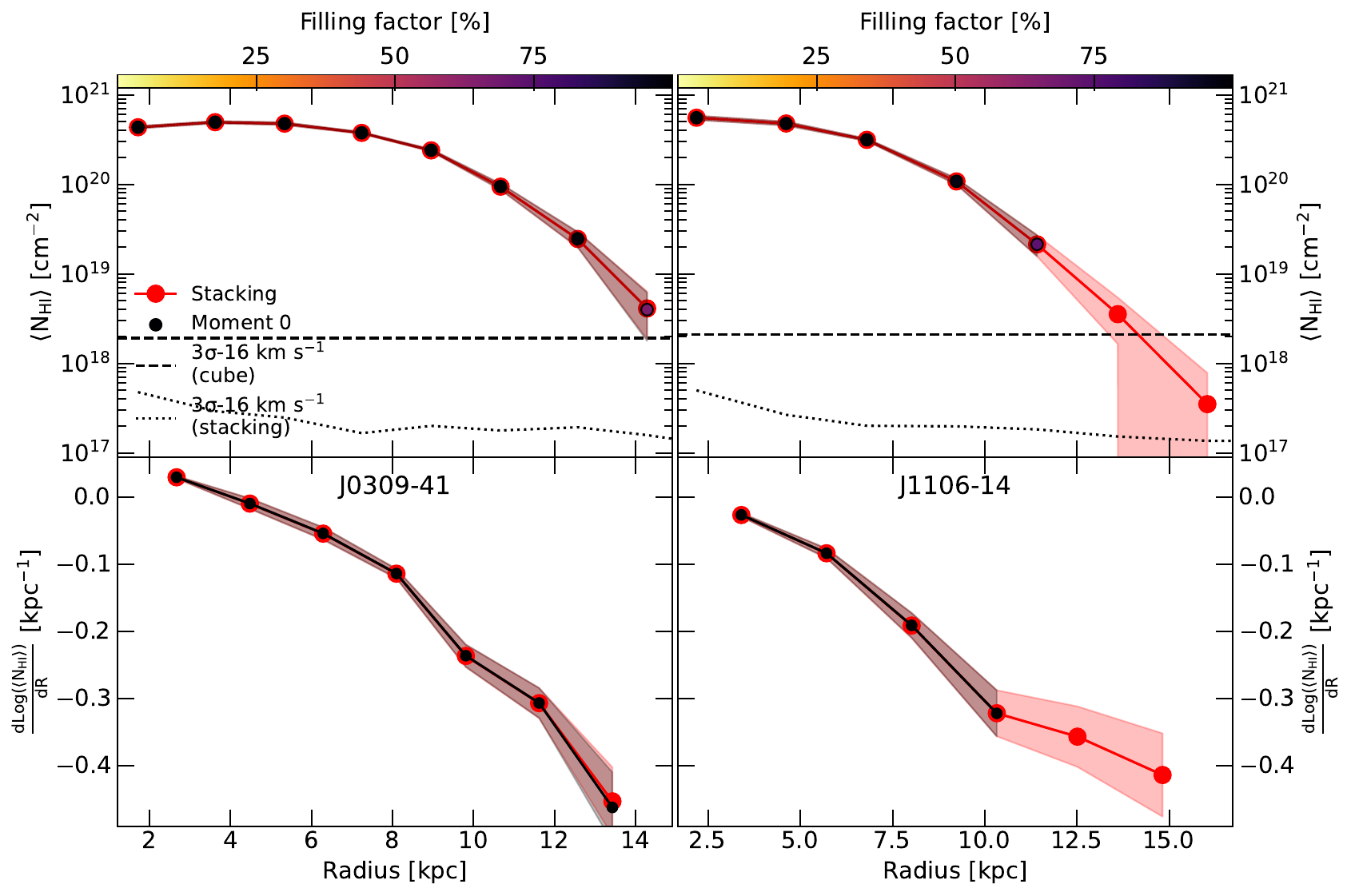}
    \caption{H\,{\sc i} radial profile (top row) and its slope (bottom row) for galaxies J1106-14 (left) and J0309-41 (right) after excluding disturbed sectors (only applies to J0309-41). The H\,{\sc i} radial profile derived from stacking and from the moment 0 map is given with red and coloured points. The colour coding reflects the ring filling factor for the moment 0 map. The black and red shaded areas are the uncertainty from the moment 0 and stacking, respectively, derived from Monte Carlo simulations. The black dashed line is the $3\sigma$ noise level in the cube over 16 km s$^{-1}$ and the black dotted line is the noise level achieved via stacking using the same definition.}
    \label{fig:cleaned}
\end{figure}

\begin{figure}
    \centering
    \includegraphics[width=\hsize]{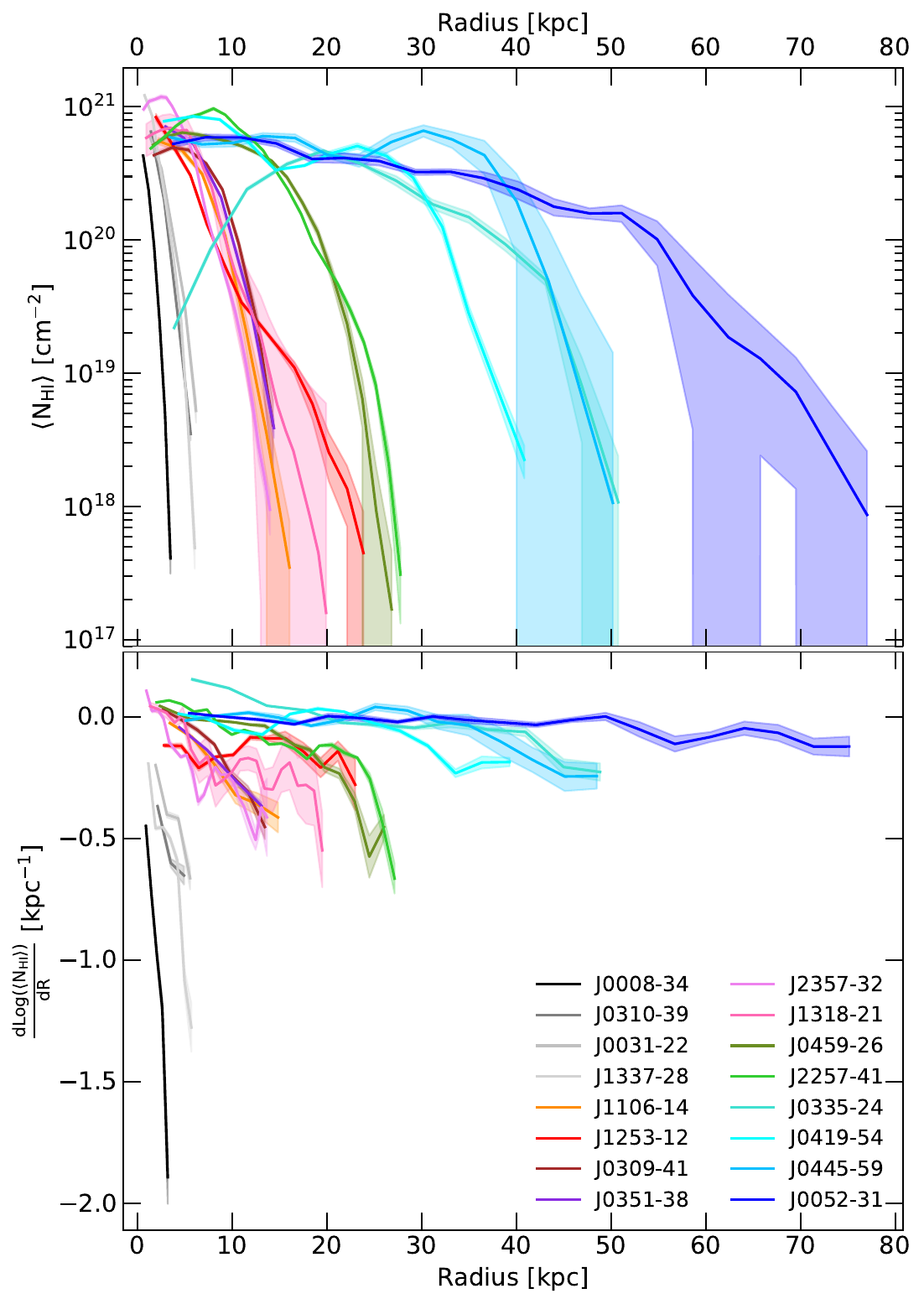}
    \caption{Clean H\,{\sc i} radial profiles (top panel) and their slope (bottom panel) on the same radial scale for the 16 MHONGOOSE galaxies studied in this work. The coloured shaded areas are the uncertainty derived from Monte Carlo simulations.}
    \label{fig:profiles}
\end{figure}

\subsection{Results}\label{sec:profile}
Having evaluated in Sect.\ \ref{sec:stacktest} the reliability of our method to recover the H\,{\sc i} radial profile of a galaxy, we applied it to the 16 MHONGOOSE galaxies considered in this study. Although we selected our sample to exclude major disturbances in the disc, a few minor perturbations such as tails and streams exist for some galaxies. To avoid such features affecting the profile, we further divide each ring into twelve $30^\circ$ wide sectors, as shown in the left panel of Fig.\ \ref{fig:rings} for the galaxy J0309-41. We derived again the H\,{\sc i} radial profile, this time after excluding the sectors that show the greatest disturbances, as shown in the panel on the right side of Fig.\ \ref{fig:rings}. To take care of the effect of the filling factor of the rings, discussed at the end of Sect.\ \ref{sec:stacktest}, we considered the H\,{\sc i} radial profile derived from the moment 0 map only for rings with filling factor $>50\%$.
\\\indent In Fig.\ \ref{fig:sectors} we provide the global and per-sector H\,{\sc i} radial profiles for J1106-14 and J0309-41 (for all our galaxies in Fig.\ \ref{fig:all_sectors}), together with their slope, which we define as
\begin{equation}\label{eq:deriv}
    \frac{\Delta {\rm Log}N_{\rm HI}}{\Delta R}=\frac{{\rm Log}N_{\rm HI}(R+\Delta R)-{\rm Log}N_{\rm HI}(R)}{\Delta R}
.\end{equation}
The figure clearly highlights those sectors which deviate from the rest of the profiles. For example, J0309-41 has a disc disturbance that distorts the profile at azimuthal angles $210^\circ<\theta<300^\circ$. In Table \ref{table:sector} we list the azimuthal sectors that were excluded for each galaxy. Deriving again the H\,{\sc i} radial profile without such regions results in the curves presented in Fig.\ \ref{fig:cleaned} for J1106-14 and J0309-41 and in Fig.\ \ref{fig:all_cleaned} for the entire sample. In Fig.\ \ref{fig:profiles} we show this `clean' H\,{\sc i} radial profile (top panel) and its slope (bottom panel) on the same scale for all galaxies. From Fig.\ \ref{fig:profiles} emerges that profiles are rather smooth. However, there seem to be two major differences between those of low-mass ($M_{\rm{HI}}\lesssim10^{9}$ M$_\odot$) and those of high-mass galaxies: the latter are less steep, as can be appreciated by looking at the slopes. The same trend is also observed in the H\,{\sc i} radial profiles presented by \citet{ianjam18} and \citet{wang25}.

\section{Discussion}\label{sec:disc}

\subsection{The presence of disc edges}\label{sec:theo}
\begin{figure}
    \centering
    \includegraphics[width=\hsize]{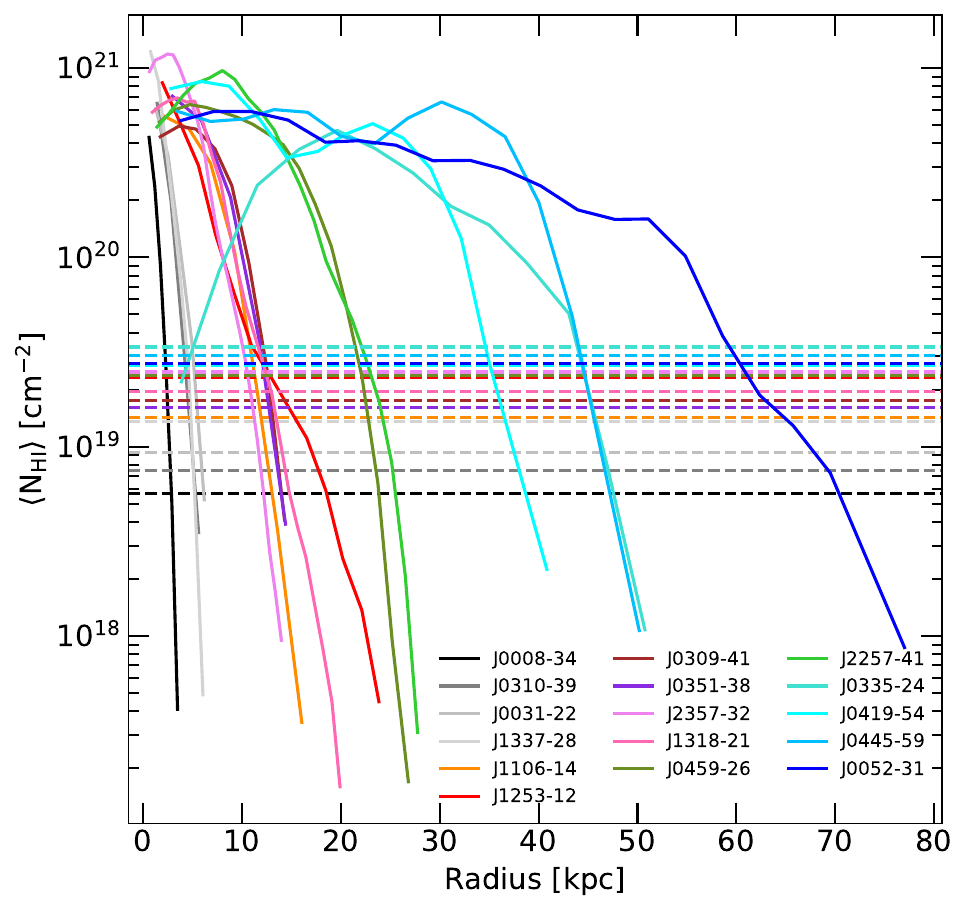}
    \caption{H\,{\sc i} radial profile of our galaxies and their column density at which the hydrogen should become mostly ionised. This critical value is reported, with the same colour of the galaxy, with the horizontal dashed lines. The colour-coding of the galaxies is the same as Fig.\ \ref{fig:profiles}. Given the crowdedness of the plot, refer also to Table \ref{table:critical} for the critical column density of each galaxy.}
    \label{fig:critical}
\end{figure}

The goal of this paper is to identify edges in the H\,{\sc i} discs and relate them to a physical break or photoionisation. Looking at the profiles of Fig.\ \ref{fig:profiles}, we observe knees occurring at column densities of about $\sim5\times10^{20}$ cm$^{-2}$, reflecting the column density threshold above which the formation of molecular hydrogen \citep{savage77,bigiel08} and cold ($\sim100$ K) H\,{\sc i} \citep{kanekar11} is favourable\footnote{Follow-up studies of the molecular radial profile may test whether the knee is really caused by the molecular-to-atomic transition.}, but no evident breaks. The majority of the slopes are almost monotonically decreasing in radius, as shown by the bottom panel of that figure, and the spread in their values renders any quantitative determination for the presence of breaks challenging. For example, it is not possible to identify edges at a single slope value.
\\\indent If the shape of the H\,{\sc i} radial profile is the result of the hydrogen becoming mainly ionised, then from photoionisation models we expect a break to occur at a certain critical column density \citep{maloney93,dove94,bland17}
\begin{equation}\label{eq:nc}
    N_{\rm C}=2.6\times10^{19}\left(\frac{\phi_{\rm i}\sigma_{\rm g}v}{\Sigma_{\rm h}}\right)^\frac{1}{2}{\rm cm}^{-2}
,\end{equation}
where $\phi_{\rm i}$ is the ionising background flux in units of $10^4$ phot cm$^{-2}$ s$^{-1}$, $\sigma_{\rm g}$ is the vertical velocity dispersion of the gas in units of 10 km s$^{-1}$, $v$ is the asymptotic rotation velocity of the disc in units of 150 km s$^{-1}$ , and $\Sigma_{\rm h}$ is the mass surface density of the dark matter halo in units of 100 M$_\odot$ pc$^{-2}$. This critical value is expected to be $\sim10^{19}$ cm$^{-2}$ if the atomic gas is warm ($T\sim10^4$ K) and smoothly distributed in the galactic potential. Can we identify a break at this critical column density?
\\\indent We calculated $N_{\rm C}$ for our galaxies taking the relevant parameters from the kinematical modelling (see Table \ref{table:critical} for the list of values). We used the mean rotation velocity in the outer disc as an estimator for $v$ and the mean dispersion for $\sigma_{\rm g}$, which was assumed to not vary with inclination. These simplifications have a minimal impact on the derived $N_{\rm C}$. In fact, an error of a factor of 10 in $v$ or $\sigma_{\rm g}$ will change $N_{\rm C}$ by only a factor $\sim$3. For $\phi_{\rm i}$ and $\Sigma_{\rm h}$, we opt for $\phi_{\rm i}=10^4$ phot cm$^{-2}$ s$^{-1}$ \citep{bland17} and a typical $\Sigma_{\rm h}=100$ M$_\odot$ pc$^{-2}$ \citep{saburova14}.
\\\indent In Fig.\ \ref{fig:critical} we show the H\,{\sc i} radial profiles of these galaxies together with their predicted critical column density. From the plot, we do not observe a correlation between $N_{\rm C}$ and the shape of the profiles, let alone any evidence for a break. For instance, J0052-31, J0419-54 and J2257-41 have almost the same $N_{\rm C}\sim2.7\times10^{19}$ cm$^{-2}$, but their profiles differ, with J2257-41 having a much shorter plateau in the inner part and a steeper decline. Also, J0310-39 and J0031-22 should have a break around $N_{\rm C}\sim0.9\times10^{19}$ cm$^{-2}$, but their H\,{\sc i} distribution is decreasing steeply from the very centre with no evidence for an edge at the expected $N_{\rm C}$. The same holds for the slopes at the critical column density, ranging from $-0.1$ for the most massive J0052-31 to $-1.5$ for the less massive J0008-34. Consequently, we argue that in our galaxies we are not observing the break predicted by photoionisation. Nonetheless, there is also another consideration to do: the available photoionisation models have been calibrated mostly on the \citet{gorkom91} observations of NGC 3198. Therefore, it could be that these predictions are galaxy-dependent.

\begin{table}
    \caption{Critical column density and parameters used in its computation.}
    \label{table:critical}
    \centering
    \begin{tabular}{c | c c | c}
    \hline\hline
        Galaxy & $v$ & $\sigma_{\rm g}$ & $N_{\rm C}$ \\
        ~      & km s$^{-1}$ & km s$^{-1}$ & $10^{19}$ cm$^{-2}$ \\
        \hline
        J0008-34 & 8.5 & 8  & 0.6 \\
        J0031-22 & 21  & 9  & 0.9 \\
        J0052-31 & 158 & 11 & 2.7 \\
        J0309-41 & 73  & 9  & 1.8 \\
        J0310-39 & 13  & 10 & 0.8 \\
        J0335-24 & 256 & 10 & 3.4 \\
        J0351-24 & 50  & 11 & 1.6 \\
        J0419-54 & 149 & 11 & 2.7 \\
        J0445-59 & 154 & 13 & 3.0 \\
        J0459-26 & 114 & 11 & 2.4 \\
        J1106-14 & 57  & 8  & 1.4 \\
        J1253-12 & 105 & 11 & 2.3 \\
        J1318-21 & 86  & 10 & 2.0 \\
        J1337-28 & 46  & 9  & 1.3 \\
        J2257-41 & 160 & 10 & 2.7 \\
        J2357-32 & 125 & 11 & 2.5 \\
    \end{tabular}
    \tablefoot{The values for $v$ and $\sigma_{\rm g}$ are taken from the best-fit of \texttt{3D-Barolo}. $\phi_{\rm i}=10^4$ phot cm$^{-2}$ s$^{-1}$ and $\Sigma_{\rm h}=100$ M$_\odot$ pc$^{-2}$ are assumed from \citet{bland17} and \citet{saburova14}, respectively.}
\end{table}

\begin{figure*}
    \centering
    \includegraphics[width=\hsize]{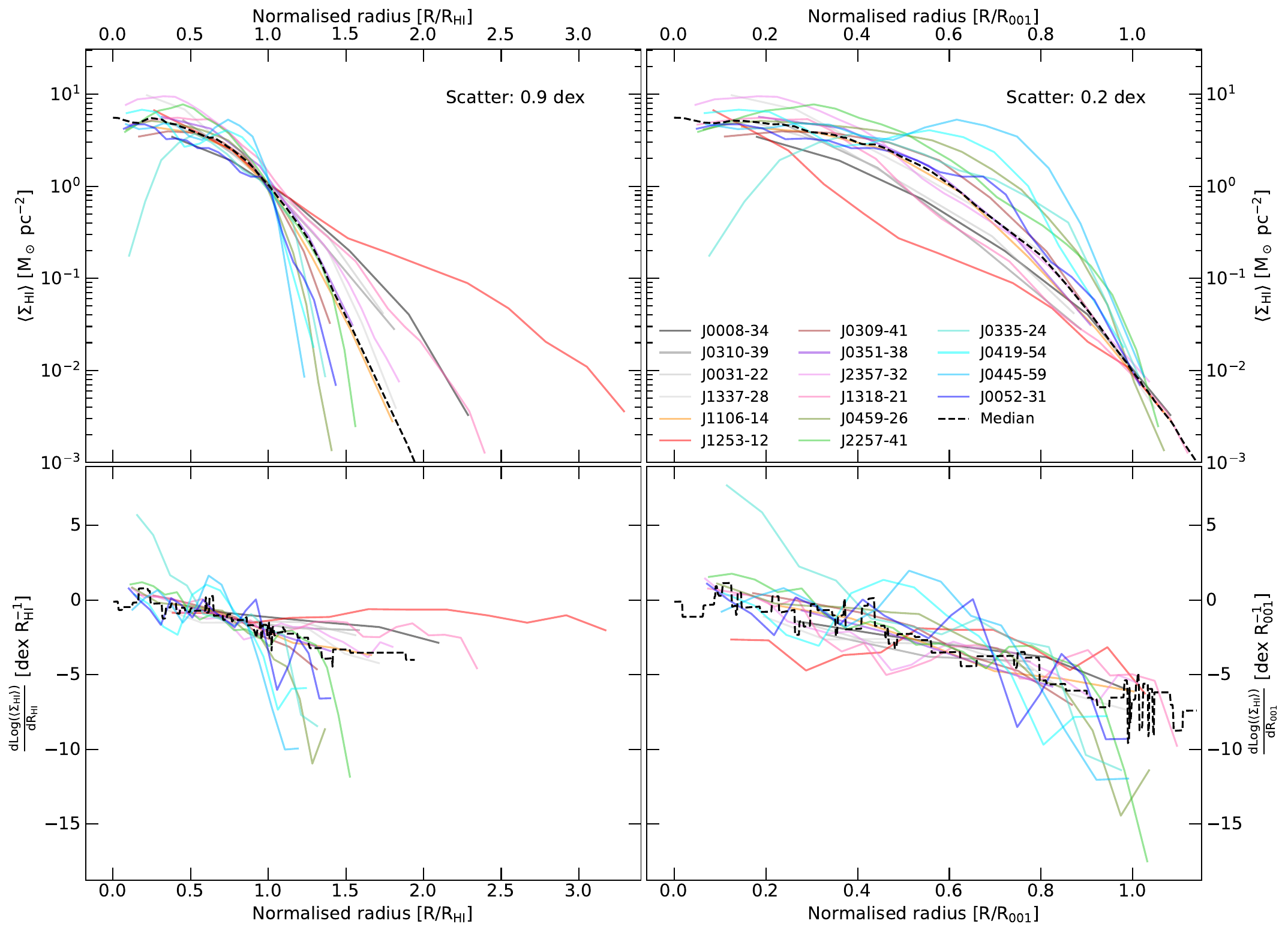}
    \caption{H\,{\sc i} radial profiles (top panels) and their slopes (bottom panels) when normalised to the radius $R_{HI}$ at which the mass surface density reaches 1 M$_\odot$ pc$^{-2}$ (left panels) and to the radius $R_{001}$ at which the mass surface density reaches 0.01 M$_\odot$ pc$^{-2}$ (right panels). The dashed curves refer to the median profile.}
    \label{fig:norm}
\end{figure*}

\begin{figure}
    \centering
    \includegraphics[width=\hsize]{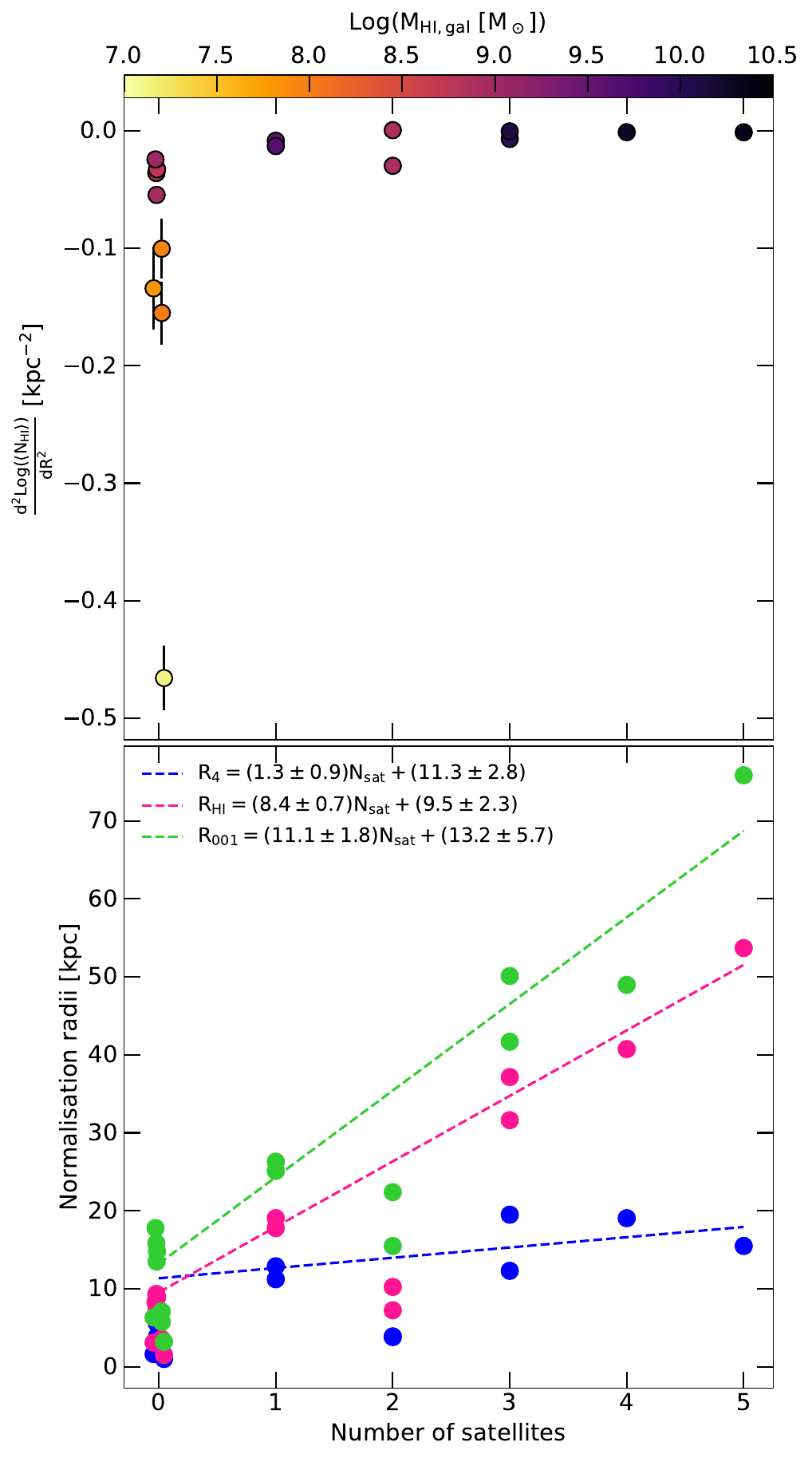}
    \caption{Slope decline (top panel) and normalisation radii (bottom panel) as a function of the number of satellites. In the top panel each circle is a galaxy, colour-coded by its H\,{\sc i} mass. Where not visible, uncertainties are smaller than the circle. In the bottom panel the blue, pink and green points refer to $R_{001}$, $R_{\rm HI}$ and $R_{4}$, respectively. The dashed lines, colour-coded with the normalisation radius they refer to, are linear least square fits considering at least one satellite. Slopes and intercepts are given in the top-left corner. In all panels a small horizontal offset has been applied to galaxies without satellites for better readability.}
    \label{fig:satellites}
\end{figure}

\subsection{What drives the shape of radial profiles?}\label{sec:mass}
The lack of evident edges in the H\,{\sc i} radial profiles and the different slopes reveal that photoionisation is likely not the main driver of the radial distribution of neutral gas. Since from Fig.\ \ref{fig:profiles} we observe that low-mass galaxies have a shorter central plateau and a steeper decline than high-mass galaxies, a first guess is that we see a resolution effect. Indeed, even if galaxies have been observed at roughly the same angular resolution, their different distances and sizes imply varying linear resolution. Therefore, it could be that the steeper slope observed for low-mass galaxies is due to the lower linear resolution with which we are sampling their radial profile. We test this by computing for each galaxy the ratio between the beam size (which is equivalent to the ring width) and the extent of the H\,{\sc i} disc. We found that all our galaxies are well-resolved, with J0310-39 having the smallest beam-to-size ratio (0.25, i.e., 4 resolution elements) and J0052-31 being the most resolved (beam-to-size ratio of 0.05, equivalent to 20 resolution elements). Consequently, resolution does not play a major role in this work.
\\\indent The observed trends between the slope of the H\,{\sc i} radial profiles and the galaxy mass might be reflecting the well-known H\,{\sc i} mass-size relation \citep{broeils97,swaters02,wang16,rajo22}, which has recently been found to hold down to column densities of $\sim10^{18}$ cm$^{-2}$ \citep{wang25}. As discussed in detail by \citet{wang16}, one interpretation for the existence of the H\,{\sc i} mass-size relation is that the physical processes acting on the H\,{\sc i} such as gas accretion \citep{faucher23}, neutral-to-molecular gas conversion \citep{saintonge22}, photoionisation \citep{madau15} and galaxy interactions \citep{ellison15}, have the same impact among different galaxies \citep{wang16}. If that picture is true, then different galaxies should exhibit H\,{\sc i} radial profiles with similar shapes and slopes when normalised to a given radius.
\\\indent In Fig.\ \ref{fig:norm} we show the H\,{\sc i} radial profiles and their slopes when normalised to $R_{\rm HI}$ (left panels), which is the radius at which a mass surface density of 1 M$_\odot$ pc$^{-2}$ is reached, and when normalised to $R_{001}$ (right panels), corresponding to the radius where the mass surface density is 0.01 M$_\odot$ pc$^{-2}$. When normalising to $R_{\rm HI}$, most\footnote{The only exception is J0335-24. This galaxy hosts a low power AGN which might be responsible for the lack of H\,{\sc i} in the inner regions of the galaxy \citep{veronese25b}.} galaxies resemble the median profile out to $R_{\rm HI}$, after which they significantly diverge (scatter of $\sim0.9$ dex). Instead, when normalising to $R_{001}$ the profiles align more closely with the median distribution (scatter of $\sim0.2$ dex). The same effect was observed by \citet{wang25}, although we must note that their sample does not include low-mass galaxies and care must be taken when comparing with their result. They interpret this behaviour as a possible hint that the physical processes acting on the low-column density H\,{\sc i} (gravity, photoionisation, gas accretion) are common to all galaxies. In contrast, since different shapes are observed when normalising to higher $\Sigma_{\rm{HI}}$ thresholds, the processes affecting the denser H\,{\sc i} distribution (mainly stellar and AGN feedback) are different for different galaxies. Our analysis seems to support this view, but with one important caveat: it holds if environmental effects are negligible.
\\\indent Among the aforementioned external processes, the environment is of particular concern. It is known from both simulations \citep{nickerson13,nierenberg13,engler21} and observations \citep{nierenberg12,sales13,geha17,mao21,mao24} that low- and high-mass galaxies generally have different environments, the latter having more satellites because of their deeper potential. The presence of companions might influence the distribution of gas in the host galaxy, for instance, by gravitationally disturbing its outer disc. Our sample does not include strongly interacting galaxies and in our analysis we already excluded the most disturbed regions of the disc (see the discussion in Sect.\ \ref{sec:profile}). However, it could still be that weak gravitational interactions produce not only clear morphological disturbances, but also more subtle effects, such as the shallower decline observed for the high-mass galaxies. To test this hypothesis, we performed a least squares linear fit to the slopes of the unnormalised radial profiles, as from the bottom panel of Fig.\ \ref{fig:profiles} it is evident how the slope values decrease approximately linearly in radius. In the top panel of Fig.\ \ref{fig:satellites}, we show this `rate of decline' for the slopes, i.e., the second derivative of the H\,{\sc i} radial profile, as a function of the number of satellites reported in Table 7 of \citet{mhongoose2}. We found that the H\,{\sc i} radial profile of galaxies with at least one companion has a second derivative close to 0, while the profile of galaxies without satellites has a higher (in absolute value) rate of decline. Furthermore, when no satellites are present, it seems that the smaller the H\,{\sc i} mass of the galaxy, the faster the rate of decline.
\\\indent Another diagnostic of the impact of satellites on the H\,{\sc i} radial profiles comes from checking whether there is a correlation between environment and normalisation radii. Indeed, the gravitational interaction likely does one (or more) of the following: redistribute the gas outward via tidal torques, build extended low-density envelopes, and prevent sharp truncations. This effectively increases the normalisation radius, especially at lower column density thresholds. Therefore, we plotted three different normalisation radii as a function of the number of satellites (see the bottom panel of Fig.\ \ref{fig:satellites}): $R_{001}$, $R_{\rm HI}$ and $R_{4}$, i.e., the radius at which a mass surface density of 4 M$_\odot$ pc$^{-2}$ is reached. The latter ($N_{\rm {HI}}\sim5\times10^{20}$ cm$^{-2}$) roughly corresponds to the location of the knee in the unnormalised radial profiles (see Sect.\ \ref{sec:theo}) and traces the inner regions of the H\,{\sc i} disc where the H\,{\sc i}-to-H$_2$ transition is likely to occur \citep{savage77,bigiel08}. We did a simple least squares linear fit to galaxies with at least one companion to quantify the correlation. The outcome is that we see a stronger linear correlation towards normalisation radii probing lower column densities:
\begin{align}
    &R_{4}=(1.3\pm0.9)N_{\rm sat}+(11.3\pm2.8)\nonumber\\
    &R_{HI}=(8.4\pm0.7)N_{\rm sat}+(9.5\pm2.3)\\
    &R_{001}=(11.1\pm1.8)N_{\rm sat}+(13.2\pm5.7)\nonumber\\
,\end{align}
where $N_{\rm{sat}}$ is the number of satellites. This means that even gentle interactions impact the H\,{\sc i} distribution. Quantifying the impact of the environment, for instance, by considering the total mass of the satellites or by including disturbed regions of the disc, as well as interacting galaxies in our sample, requires a larger sample size and is outside the scope of this paper, and we leave it for future investigations.
\\\indent In conclusion, we found that the radial H\,{\sc i} distribution of non-interacting galaxies becomes more self-similar at progressively lower column density thresholds, in agreement with previous studies such as \citet{wang25}. In addition to what was proposed in the literature, we argue that the environment must be taken into account when interpreting this self-similarity. Indeed, the presence of satellites, even if they are not causing direct gas stripping, redistributes the gas to larger radii, producing extended, shallow outer profiles. When scaled by an environmentally dependent radius, these profiles appear self-similar, but this similarity arises from the scaling itself rather than from universal underlying physics.

\section{Conclusion and future prospect}\label{sec:conc}
We presented the H\,{\sc i} radial profiles of 16 MHONGOOSE galaxies. We probed, at spatial resolutions from 0.6 kpc to 4 kpc depending on galaxy distance, the gaseous disc down to column densities of a few times 10$^{17}$ cm$^{-2}$, more than an order of magnitude lower than the expected column density at which the disc should turn mostly ionised \citep{maloney93,dove94,bland17} and an order of magnitude lower than a previous observational analysis based on THINGS and HALOGAS observations \citep{ianjam18}. In the following, we list our main findings:
\begin{enumerate}[i)]
    \item The H\,{\sc i} radial profiles are characterised by an inner plateau, likely associated with the transition from atomic to molecular hydrogen, followed by a knee at column densities of about $5\times10^{20}$ cm$^{-2}$. We do not observe breaks at column densities of few times $10^{19}$ cm$^{-2}$ at which the hydrogen should become mostly ionised, as predicted by photoionisation models.
    \item We found that the H\,{\sc i} radial profiles are self-similar when normalised to the radius at which a mass surface density of 0.01 M$_\odot$ pc$^{-2}$ is reached, in agreement with recent work by \citet{wang25}.
    \item We also found that galaxies with at least one satellite do not show a steep decline, and that the normalisation radius correlates with the number of satellites, suggesting that the environment plays a role in shaping the H\,{\sc i} distribution in galaxies.
\end{enumerate}
Our analysis was limited by mainly two observational factors: sample size and environment. In future, insights on what drives the H\,{\sc i} distribution in galaxies might come from a more precise quantification for the impact of environment. Considering a statistically significant number of low-mass galaxies, which are expected to not retain satellites, might constitute a good control sample to test against while simultaneously confirm whether it is true that they have a steeper profile than high-mass systems.
\\\indent The thorough revision of photoionisation models may also be necessary. For instance, Eq.\ (\ref{eq:nc}) assumes that the H\,{\sc i} velocity dispersion outside the stellar disc is set by the gravitational potential of the dark matter halo, i.e., the volume density is dominated by dark matter. However, recent studies found that the volumetric densities of gas in these regions could be much larger (e.g., \citealt{bacchini19}) and it may significantly contribute to the gravitational force perpendicular to the disc, affecting the velocity dispersion and, consequently, the analytical form of Eq. (\ref{eq:nc}). New results from the eROSITA X-ray telescope \citep{erosita} confirm the presence of a hot CGM \citep{zhang24} whose X-ray luminosity correlates with the galaxy stellar mass \citep{zhang24b}. The pressure from the hot medium may compress the H\,{\sc i} disc, decreasing its thickness and increasing the gas volumetric density \citep{wang24b,lin25}. Extending then the revised photoionisation models to a larger sample of galaxies, can give invaluable information for a better estimate of the CIB intensity and the ionisation state of the CGM, impacting our understanding of galaxy formation and evolution.

\begin{acknowledgements}
We thank the anonymous referee for the constructive comments. This work has received funding from the European Research Council (ERC) under the European Union’s Horizon 2020 research and innovation programme (grant agreement No 882793 `MeerGas'). The MeerKAT telescope is operated by the South African Radio Astronomy Observatory, which is a facility of the National Research Foundation, an agency of the Department of Science and Innovation. This research has made use of the NASA/IPAC Extragalactic Database (NED), which is funded by the National Aeronautics and Space Administration and operated by the California Institute of Technology. JH acknowledges support from the UK SKA Regional Centre (UKSRC). The UKSRC is a collaboration between the University of Cambridge, University of Edinburgh, Durham University, University of Hertfordshire, University of Manchester, University College London, and the UKRI Science and Technology Facilities Council (STFC) Scientific Computing at RAL. The UKSRC is supported by funding from the UKRI STFC.
\end{acknowledgements}

\bibliographystyle{aa}
\bibliography{reference.bib}

\appendix
\section{Testing the method}\label{sec:test}
\subsection{Uncertainty estimation}\label{sec:error}
The uncertainty associated with the radial profile includes a number of different effects, such as the angular resolution, the uncertainty on the parameters of the kinematical modelling, and assumptions on disc geometry and kinematics outside the detection mask. Quantifying the impact of each source of uncertainty is non-trivial.

\subsubsection{Resolution}
\begin{figure}
    \centering
    \includegraphics[width=\hsize]{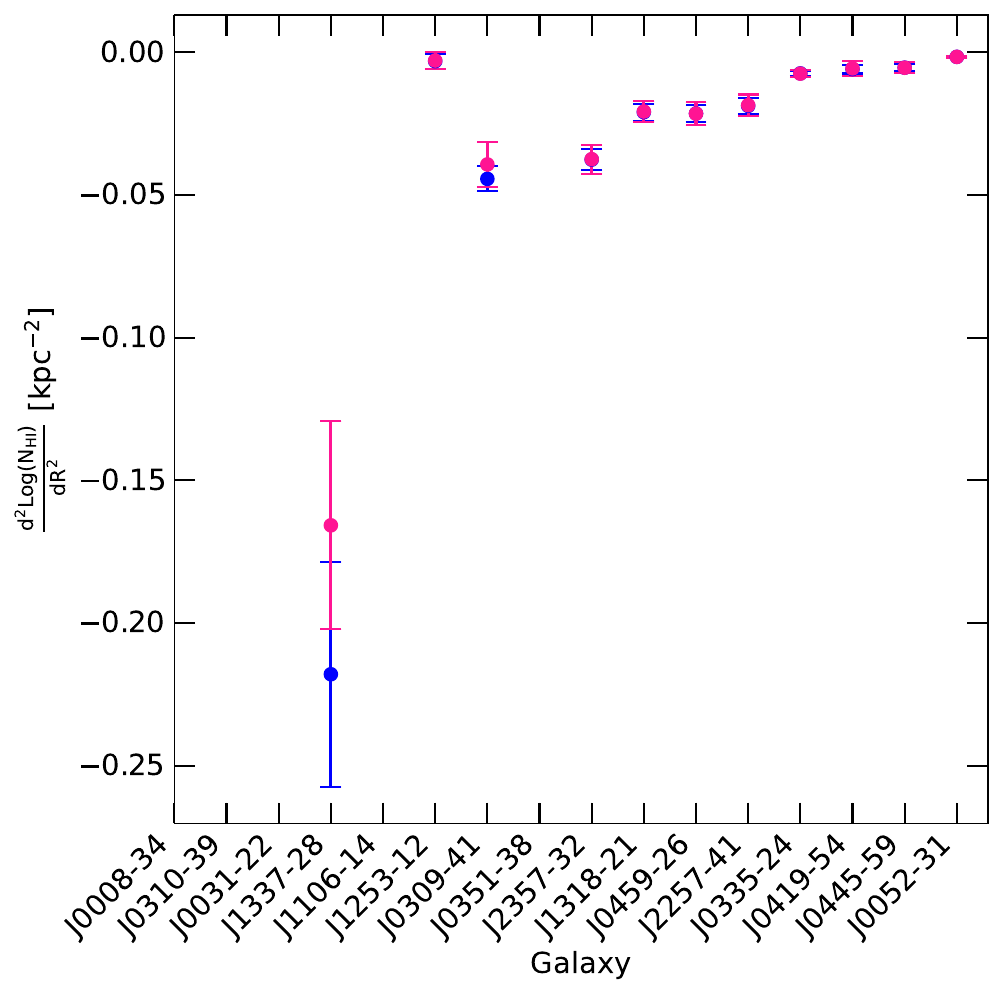}
    \caption{Comparison between the second derivative of the H\,{\sc i} radial profiles at native resolution (blue) with the second derivative of the H\,{\sc i} radial profiles undersampled by a factor of 2 (pink). No data are shown for galaxies J0008-34, J0310-39 and J0031-22 as we do not have enough points in their undersampled profile to compute the second derivative.}
    \label{fig:reserror}
\end{figure}

MHONGOOSE galaxies are well resolved, the least resolved being J0310-39 with four resolution elements. However, the finite angular resolution implies that computing the slope using Eq.\ \ref{eq:deriv} might be sensitive to how large $\Delta R$ is. Indeed, one would expect that the larger $\Delta R$, the larger the deviation from the expected value of the slope, i.e., the proper derivative $d\rm{(LogN}_{H\sc i})/dR$. We estimated the impact of a finite $\Delta R$ by degrading the resolution of the profiles by a factor of 2, that is, by sampling the profiles presented in Fig.\ \ref{fig:cleaned} every two radial points. In Fig.\ \ref{fig:reserror} we compare the second derivative of the original profiles with their undersampled version. It is clear that resolution does not play a major role in our case, as the second derivatives are consistent within $1\sigma$ and their uncertainty does not change significantly when the angular resolution is degraded. 

\subsubsection{Tilted-ring fit}
The rings used to compute the H\,{\sc i} radial profiles were created by averaging the geometry of the disc as given by the 3D fitting described in Sect.\ \ref{sec:rings}. This effectively means that we are assuming the disc to have a uniform and constant geometry. To test the impact of this assumption, as well as the impact of the uncertainty on the inclination and position angle of the rings, we run for each galaxy 5000 Monte Carlo simulations. Each simulation consists of extracting the radial profile from the moment 0 map and via stacking by varying the inclination and position angle of the rings according to the uncertainties reported in Table \ref{table:angles}, and accounting for regions where tails and asymmetries are visible in the moment 0 map (as discussed in Sect.\ \ref{sec:profile}). As shown in Fig.\ \ref{fig:profiles}, the uncertainty associated with geometrical errors is relevant only for galaxies J0052-31, J0445-59 and J1318-21, meaning their disc is likely too warped to assume constant geometry. However, the conclusion about lack of an edge associated with photoionisation and about the shallower decline with respect to low-mass galaxies still holds.

\subsubsection{Geometry and kinematics assumptions}
For most galaxies we are tracing the radial profile out to only one beam outside the \texttt{SoFiA-2} detection mask, despite the employing of spectral stacking. The lack of a significant amount of H\,{\sc i} in the outer regions is consistent with the result presented by \citet{veronese25} and \citet{marasco25} based on the comparison between the observed H\,{\sc i} distribution around MHONGOOSE galaxies with the predicted distribution around Milky Way-like galaxies from cosmological simulations. Therefore, we do not expect that the velocity field extension described in Sect.\ \ref{sec:stack} significantly affects our results, as we cannot detect H\,{\sc i} emission in such regions even via stacking. Furthermore, \citet{veronese25} discussed the difficulties in choosing the correct velocity extension given our ignorance about the kinematics of the gas at such large galactocentric radii. They showed that is hardly any difference between employing a constant rotation curve, as done here, or no rotation at all. Other studies have also shown that an alignment error of up to 50 km s$^{-1}$ does not significantly impact the outcome of H\,{\sc i} spectral stacking \citep{khandai11,maddox13,neumann23}. Given the low level of emission detected outside the \texttt{SoFiA-2} mask, we argue that our assumption of constant disc geometry in these regions, i.e., no warp, also has little impact on our conclusions.

\subsection{How effective is our method?}\label{sec:stacktest}
\begin{figure*}
    \centering
    \includegraphics[width=0.93\hsize]{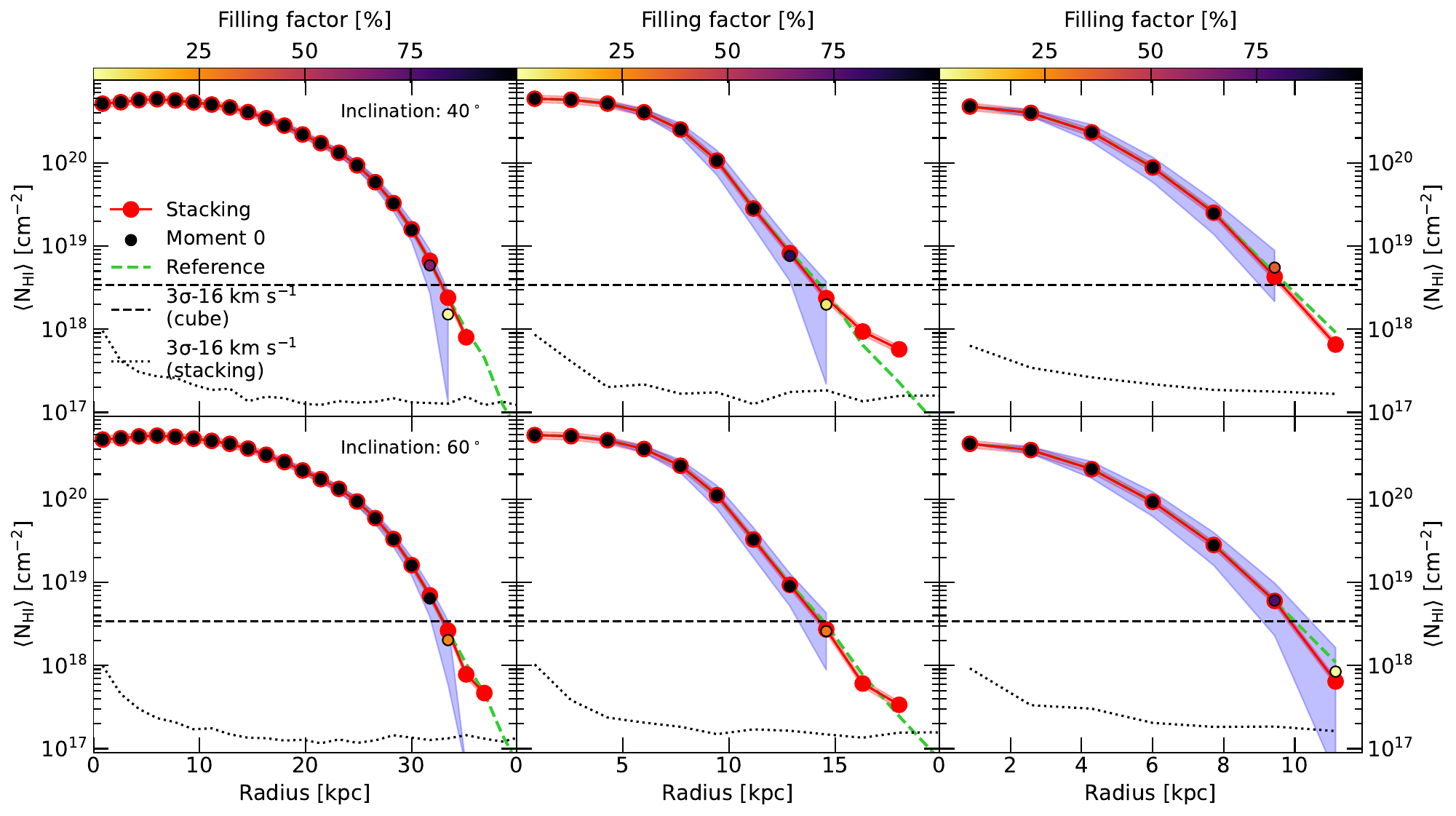}
    \caption{H\,{\sc i} radial profiles of the six mock galaxies. Galaxies in the first row have inclination of $40^\circ$ and in the second row of $60^\circ$. In all panels, the reference profile from \texttt{GALMOD} is given with a green dashed line, whereas the points, colour-coded with the filling factor of each ring, are the radial profile derived from the moment 0 map. The red points, instead, are the radial profile determined from the stacked spectra. The blue and red shaded areas are the uncertainty from the moment 0 and stacking, respectively. The former is defined as the standard deviation of the intensity values in each ring, whereas the latter as 10\% of the total flux in the line, equivalent to the typical uncertainty on the calibration of the H\,{\sc i} flux. The black dashed line is the $3\sigma$ noise level in the cube over 16 km s$^{-1}$. The black dotted line is noise level achieved via stacking using the same definition.}
    \label{fig:mock}
\end{figure*}

We tested the ability of our method to recover the H\,{\sc i} radial profiles. For this we build a sample of six mock galaxies. Each model is created on a set of rings whose width along the major axis is $35''$. This value is equal to the average major axis of the synthesised beam of the MHONGOOSE observations used in this paper. Two galaxies have a maximum radius of $900''$, two of $450''$, and two of $200''$. We placed them at a distance of 10 Mpc, so as to achieve physical sizes ($R_{\rm max}$) of 45 kpc, 22.5 kpc and 12 kpc. These are representative of the radial extent of the MHONGOOSE galaxies studied here. For each size, one model has an inclination of 40$^\circ$ and the other of 60$^\circ$, i.e., spanning the range of inclinations in our MHONGOOSE subsample. We defined a rotation curve for each size. It rises linearly to a maximum velocity, which is reached at half of the maximum radius, and is flat afterwards. We chose maximum velocities of 115 km s$^{-1}$, 100 km s$^{-1}$ and 80 km s$^{-1}$ for the largest, intermediate, and smallest galaxies. Lastly, we defined a radial H\,{\sc i} distribution approximately constant in the inner parts followed by an exponential decline.
\\\indent We used the ring parameters and the rotation curve to build six mock H\,{\sc i} cubes via the \texttt{GALMOD} task of \texttt{3D-Barolo}, originally included in the Groningen Image Processing System (\texttt{GIPSY}; \citealt{gipsy,gipsy2}). We created a Gaussian noise cube and convolved it with the average beam of the MHONGOOSE observations ($35''\times27''$) in order to reach a noise level comparable to that of real observations ($3\sigma$ limit of $2\times10^{18}$ cm$^{-2}$ over 16 km s$^{-1}$). This noise cube was coadded to the mock galaxies. We then produced moment 0 maps and velocity fields by running \texttt{SoFiA2} with the same source finding parameters used for MHONGOOSE \citep{mhongoose2}. This leads to data products as close as possible to real MHONGOOSE observations.
\\\indent We derived the observed H\,{\sc i} profiles from the moment 0 map of each mock galaxy by computing the average column density in the rings corrected for the inclination. Blank or masked pixels are not included in the computation. We also derived the H\,{\sc i} radial profile using the stacking procedure described in Sect.\ \ref{sec:stack}. In this case, all the LoS of the annuli were considered. We measured the total flux detected in the stacked spectrum of each ring and converted the value into the face-on column density.
\\\indent Looking at the result of the test, provided in Fig.\ \ref{fig:mock}, we can evaluate the implications of our approach. For all mock galaxies, above the $3\sigma$ noise limit of the cube the H\,{\sc i} profile retrieved from the moment 0 map and from stacking are consistent with each other and with the reference profile built by \texttt{GALMOD}. The filling factor of the rings on the moment map, i.e., the number of LoS included in the \texttt{SoFiA2} mask divided by the total number of LoS in the ring, is also $>50\%$ in this regime. As soon as the filling factor drops below 50\% we cannot trace the H\,{\sc i} profile with the moment 0 map and have to rely on stacking.

\subsection{What does an edge look like?}\label{sec:edgetest}
\begin{figure*}
    \centering
    \includegraphics[width=0.93\hsize]{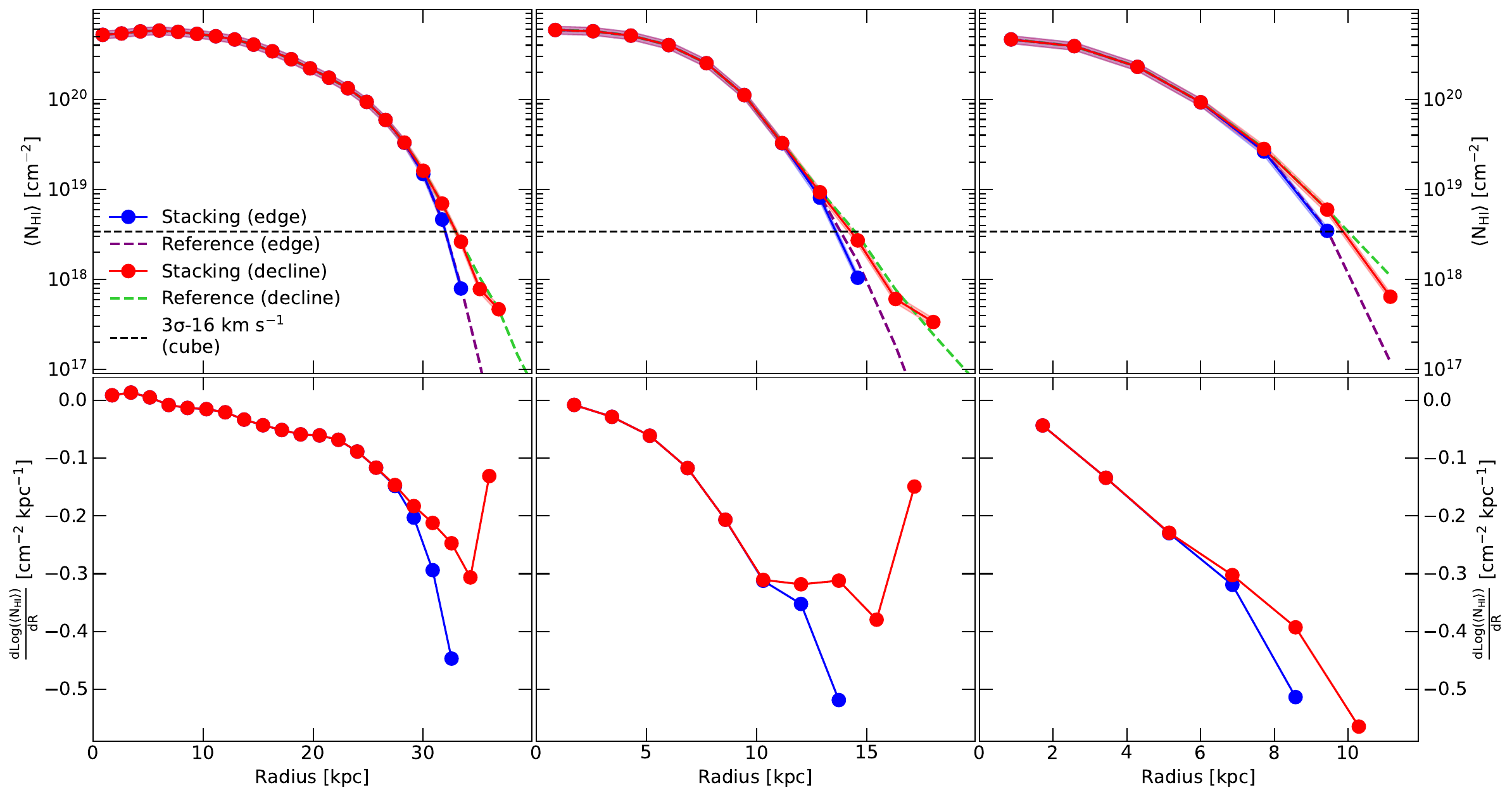}
    \caption{Comparison of the H\,{\sc i} radial profiles of three mock galaxies having a smoothly declining H\,{\sc i} distribution or a disc edge. In the top panels, the reference profile from \texttt{GALMOD} is given with a dashed line (green for the smoothly declining distribution and purple for the distribution with an edge), whereas the points (red for the smoothly declining distribution and blue for the distribution with an edge) are the radial profiles determined from the stacked spectra. The blue and red shaded areas are the uncertainties on the flux in the stacked spectra. In the bottom panels the slopes of the smoothly declining profile (red) and for the profile with an edge (blue) are shown. The black dashed line is the $3\sigma$ over 16 km s$^{-1}$ noise level in the cube.}
    \label{fig:edge}
\end{figure*}

The aim of this paper is to confirm the existence of an edge in the H\,{\sc i} radial profile of nearby galaxies as predicted by photoionisation models. We used again mock galaxies to investigate whether we can actually detect the edge given the angular resolution of the data and the approach we used. Therefore, we created another set of three mock galaxies with a sharp edge at column densities of $\sim10^{19}$ cm$^{-2}$.
\\\indent To create the edge we took the H\,{\sc i} distribution of the models with a 60$^\circ$ inclination and clipped it below the column density of 10$^{19}$ cm$^{-2}$, i.e., all values $<10^{19}$ cm$^{-2}$ are set to 0. Again we used \texttt{GALMOD} to create the mock cubes based on this input distribution and repeated the procedure to derive the H\,{\sc i} radial profiles. In Fig.\ \ref{fig:edge} we compare in the upper panels the smoothly declining profiles with the edged profiles and in the lower panels their slopes.
\\\indent This test shows that at the angular resolution of our data ($35''$) the edge is seen as a sudden change in the slope value in the last radius detected. In particular, the slope reaches a value of $\sim-0.5$ (close to the expected value of $-0.4$ from photoionisation models) no matter the size of the galaxy.

\clearpage\onecolumn
\section{Ancillary plots}\label{app:plots}
\begin{figure}[!h]
    \centering
    \includegraphics[width=.90\hsize]{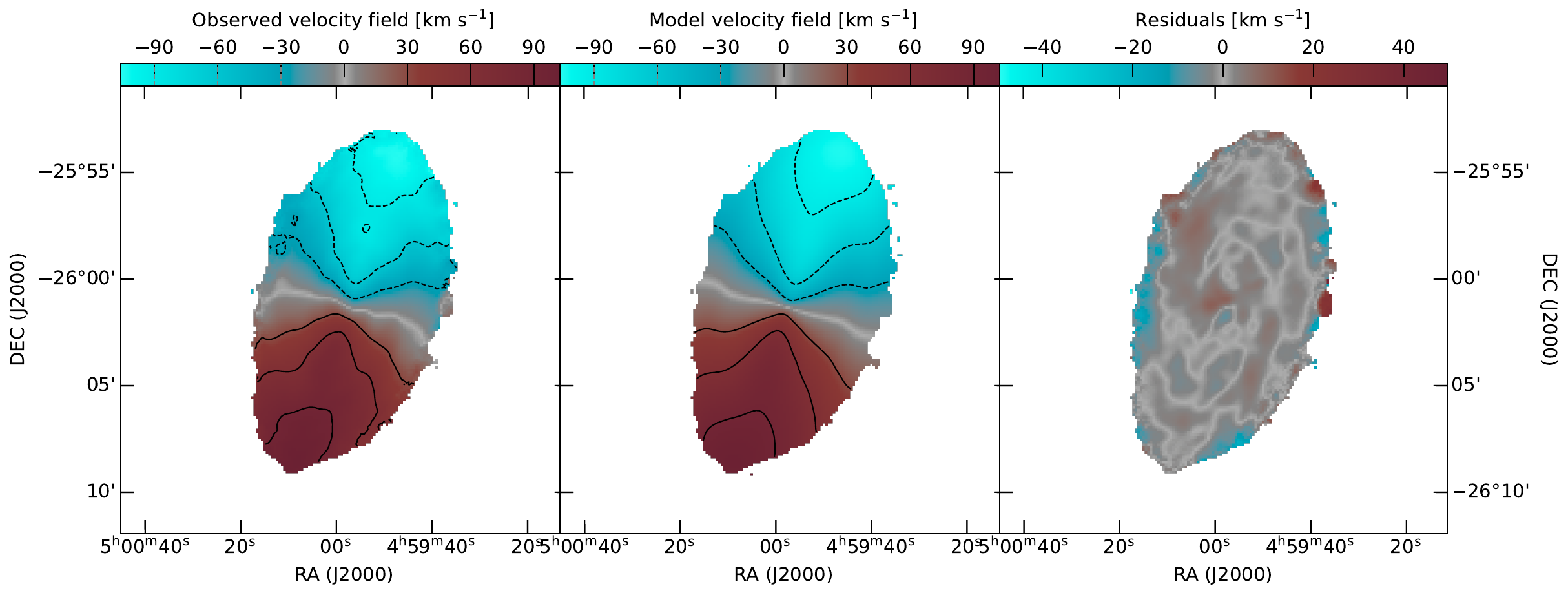}
    \caption{Comparison between J0459-26 observed and modelled velocity field. Left-to-right are shown the observed, modelled and residual velocity field. The model was derived from \texttt{3D-Barolo}. The residuals are in terms of data$-$model. In the first and second panels, dashed contours refer to the approaching side, whereas solid lines correspond to receding velocities. The contour levels are given at the colorbar on top of each plot with respect to the systemic velocity of 739.9 km s$^{-1}$.}
    \label{fig:mom1}
\end{figure}

\begin{figure}[!h]
    \centering
    \includegraphics[width=.90\hsize]{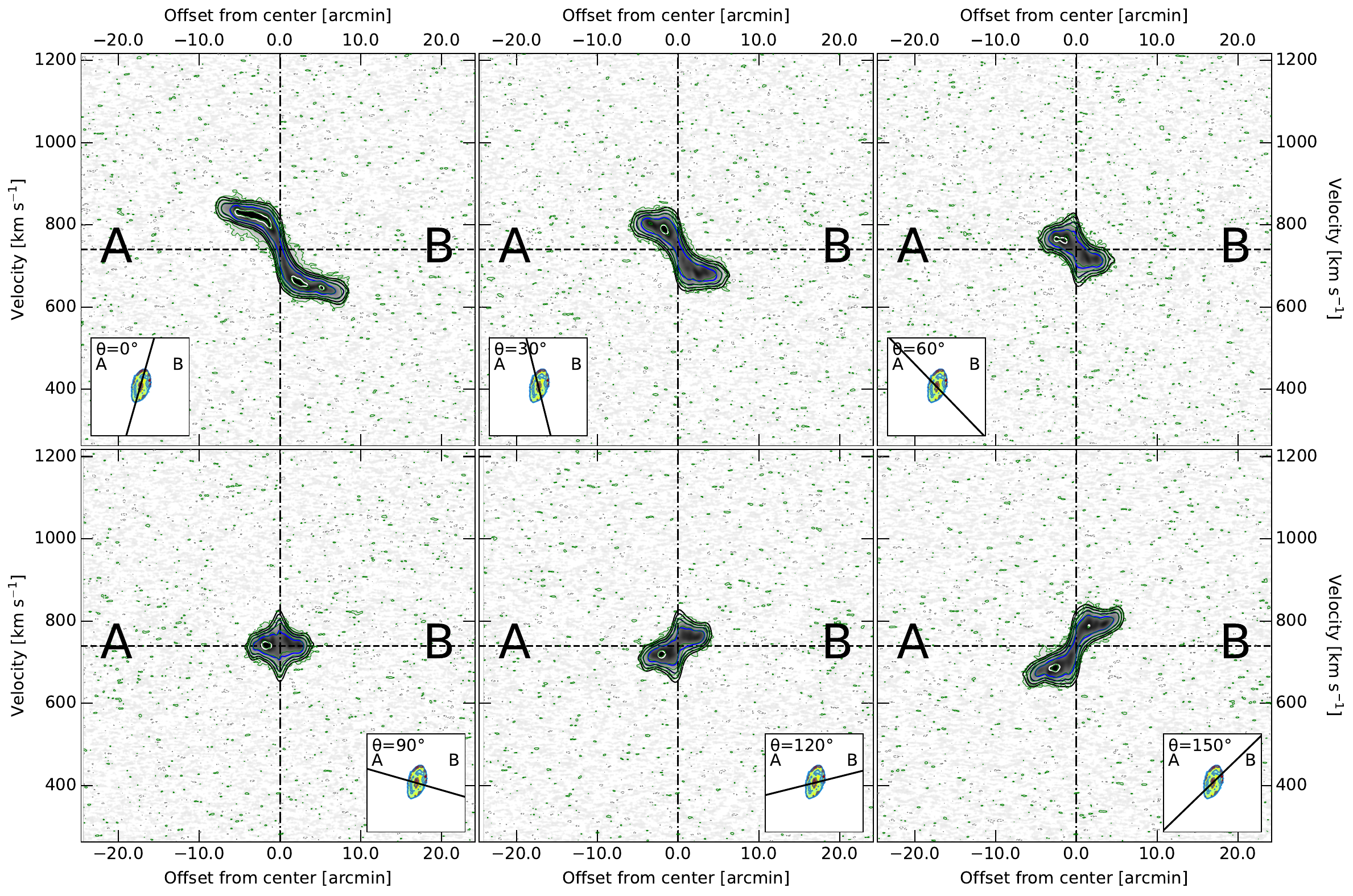}
    \caption{J0459-26 $pv$ diagrams at azimuthal angles ($\theta$) of (0, 30, 60, 90, 120, 150) degrees for the data and the best-fit model derived by \texttt{3D-Barolo}. The model was determined by fitting the approaching and receding side of the galaxy simultaneously. Both the data and the model cube were Hanning-smoothed to a velocity resolution of 7 km s$^{-1}$. For each panel, the background grey-scale image is the beam-wide slice extracted from the data cube at a given $\theta$. Solid contour levels are denoting the (4, 16, 64, 256, 1024)$\sigma$ level. The coloured lines refer to the model, the green curves to the data. The dashed grey contour refers to the $-4\sigma$ level. The dashed black horizontal line corresponds to the systemic velocity, whereas the dashed-dotted black vertical line denotes the galaxy centre. The insets show the moment 2 map and the slice along which the $pv$ is extracted. Letters $A$ and $B$ provide the orientation in the main panels.}
    \label{fig:pv}
\end{figure}

\begin{figure*}[!p]
    \centering
    \includegraphics[width=\hsize]{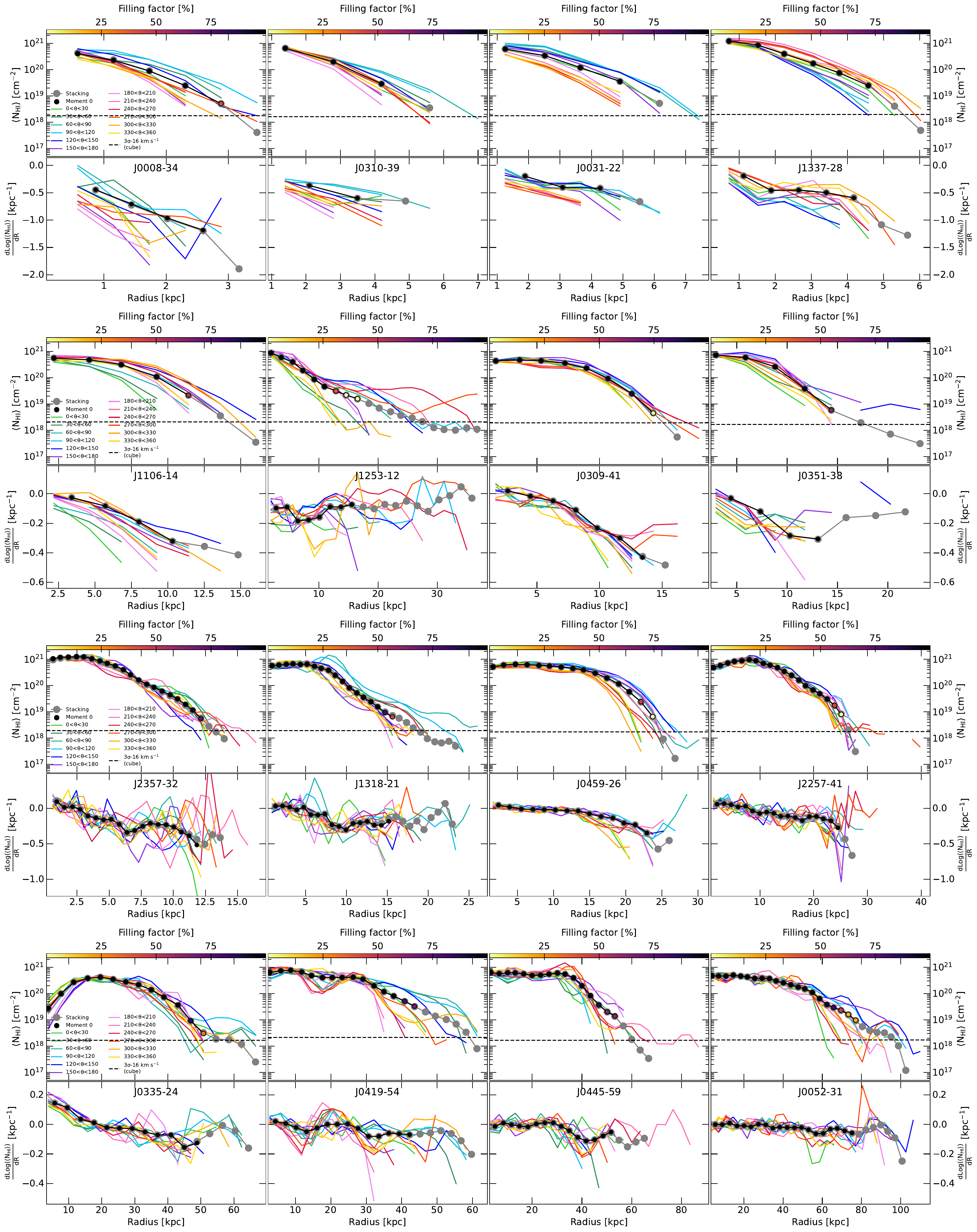}
    \caption{H\,{\sc i} radial profile (top) and its slope (bottom) for each azimuthal sector for each galaxy. The panels are left-to-right and top-to-bottom in increasing order of galaxy H\,{\sc i} mass. Panels description follows Fig. \ref{fig:sectors}.}
    \label{fig:all_sectors}
\end{figure*}

\begin{figure*}[!p]
    \centering
    \includegraphics[width=\hsize]{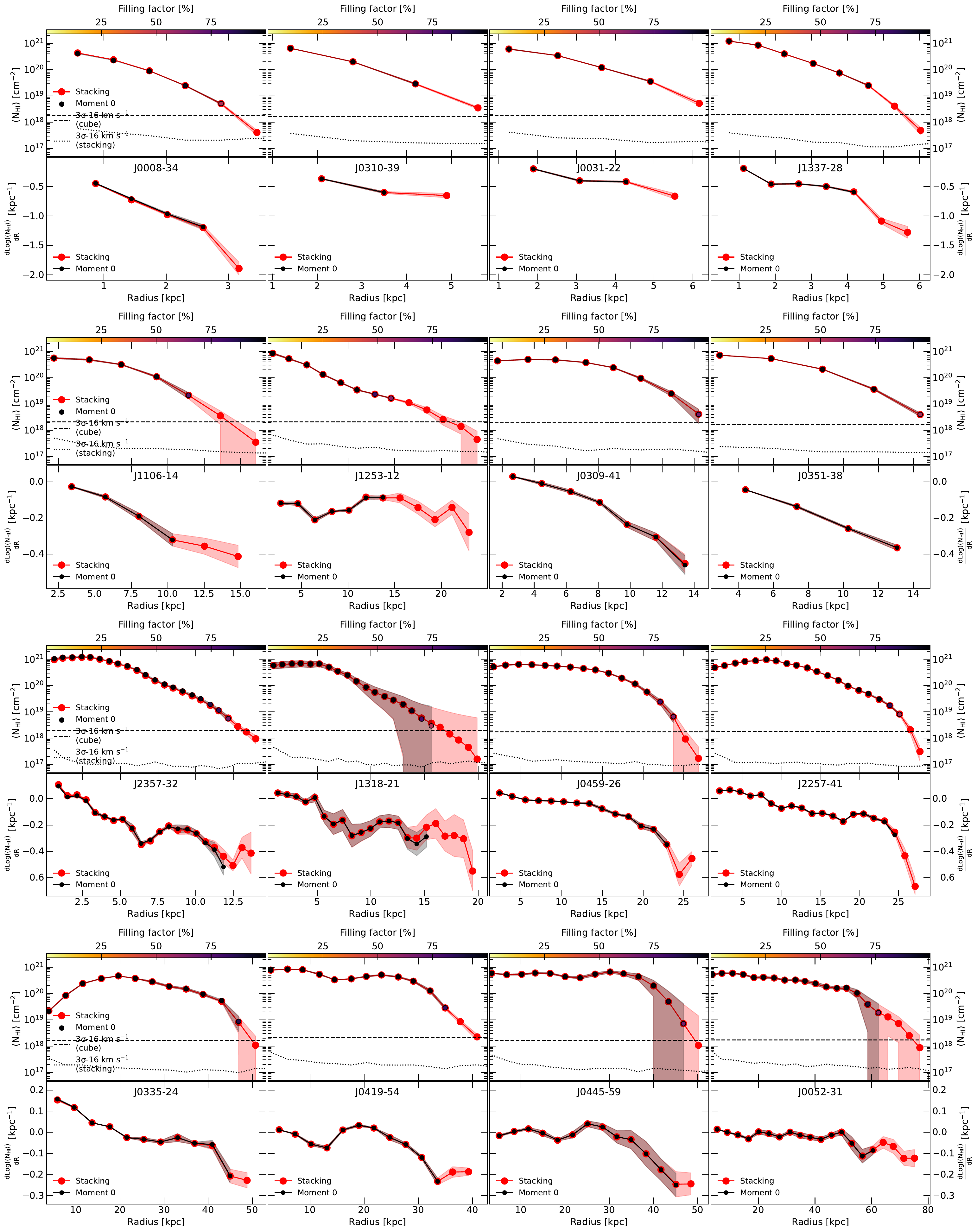}
    \caption{H\,{\sc i} radial profile (top) and its slope (bottom) for each galaxy after excluding the disturbed sectors. The panels are left-to-right and top-to-bottom in ascending order of galaxy H\,{\sc i} mass. Panels description follows Fig. \ref{fig:cleaned}.}
    \label{fig:all_cleaned}
\end{figure*}

\clearpage\twocolumn
\section{Comparison with THINGS}\label{sec:things}
\begin{figure}
    \centering
    \includegraphics[width=\hsize]{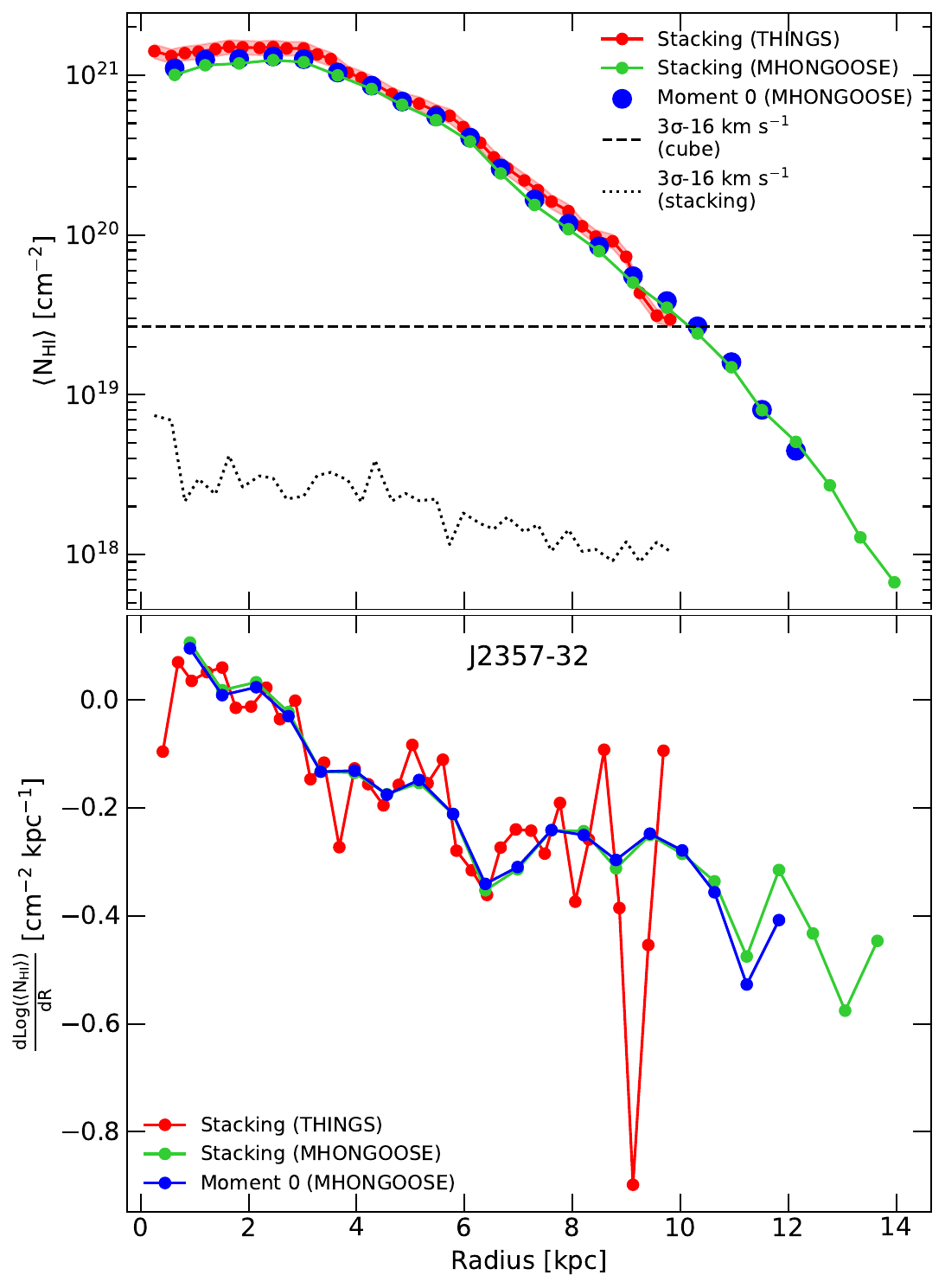}
    \caption{Comparison between the MHONGOOSE and THINGS H\,{\sc i} radial profiles of the galaxy J2357-32. In the top panel is shown the H\,{\sc i} radial profile derived from the MHONGOOSE (blue points) moment 0 map. The red and green curves are the H\,{\sc i} radial profiles derived from stacking the THINGS and the MHONGOOSE cubes, respectively. The red shaded area denote the uncertainty on the stacking points, measured as the 10\% of the line flux. This corresponds to the typical calibration error in the flux. The horizontal dashed line and the dotted curve indicate the noise level in the moment 0 map and stacked spectra of the THINGS data. In the bottom panel the radial slope of the H\,{\sc i} profile is shown for the above datasets and methods.}
    \label{fig:things}
\end{figure}

The MHONGOOSE galaxy J2357-32 (NGC 7793) is included in THINGS and its H\,{\sc i} radial profile was also presented by \citet{ianjam18}, thus allowing a direct comparison. The authors aimed at quantifying disc edges in nearby galaxies, and for this they used a combination of THINGS and HALOGAS data. Via stacking they probed the H\,{\sc i} radial profile down to $\sim10^{19}$ cm$^{-2}$ in 9 out of 17 galaxies, including J2357-32, and a few times $10^{18}$ cm$^{-2}$ for the others. They found no evidence for the predicted sharp H\,{\sc i} column density break and suggested that factors other than photoionisation, such as late-stage gas accretion from the cosmic web, may determine the extent of H\,{\sc i} discs, recommending deeper observations to confirm this. MHONGOOSE provided such observations and so we compare here the NGC 7793 radial profile between THINGS and MHONGOOSE to show the improvement in column density and to look for the presence of an edge.
\\\indent We used the THINGS natural-weighted data \citep{things} including the H\,{\sc i} cube, the primary beam corrected moment 0 map and the moment 1 map. The beam size of the data is $15.6''\times10.8''$, the spectral resolution is 2.6 km s$^{-1}$ and the $3\sigma$ noise level is $2.7\times10^{19}$ cm$^{-2}$ over 16 km s$^{-1}$. We derived the H\,{\sc i} radial profile following the same procedure used for the MHONGOOSE data, with the only difference that the width along the major axis of the elliptical annuli has been set equal to the THINGS beam size ($15.6''$). In Fig.\ \ref{fig:things} we show the outcome.
\\\indent The improvement in tracing the H\,{\sc i} radial profile provided by MHONGOOSE with respect to THINGS is evident. Indeed, MHONGOOSE extends the profile radially by 25\% just with the moment 0 map, and by 35\% when stacking is employed. These results come from the much better sensitivity, which allow to probe the profile down to column densities at least 1 dex lower than THINGS. However, despite those improvements, we cannot identify a clear disc edge in our deeper MHONGOOSE data either.

\label{LastPage}
\end{document}